\documentclass{scrartcl}
\usepackage[utf8]{inputenc}
\usepackage{authblk}
\usepackage[numbers,sort&compress]{natbib}
\usepackage{doi}
\usepackage{amsmath}
\usepackage{graphicx}
\usepackage{fancyhdr}
\usepackage{xcolor}
\usepackage{soul}
\usepackage[normalem]{ulem}

\title{Transition from no-ELM response to pellet ELM triggering during pedestal build-up -- insights from extended MHD simulations}

\author[1]{\small S Futatani}
\author[2,3]{A Cathey}
\author[2]{M Hoelzl}
\author[2]{P T Lang}
\author[4,5]{G T A Huijsmans}
\author[2]{M. Dunne}
\author[6]{JOREK Team}
\author[7]{ASDEX Upgrade Team}
\author[8]{EUROfusion MST1 Team}

\affil[1]{\footnotesize Universitat Politècnica de Catalunya, Barcelona, Spain}
\affil[2]{Max Planck Institute for Plasma Physics, Boltzmannstr. 2, 85748 Garching b. M., Germany}
\affil[3]{Physik Department, E28, TUM, 85748 Garching, Germany}
\affil[4]{CEA, IRFM, 13108 Saint-Paul-Lez-Durance, France}
\affil[5]{Eindhoven University of Technology, P.O. Box 513, 5600 MB Eindhoven, The Netherlands}
\affil[6]{refer to [M Hoelzl, G T A Huijsmans, S J P Pamela, M Becoulet, E Nardon, F J Artola, B Nkonga et al, Nuclear Fusion, in preparation] for a list of team members}
\affil[7]{see the author list of [H. Meyer et al. 2019 Nucl. Fusion 59 112014]}
\affil[8]{see the author list of [B. Labit et al. 2019 Nucl. Fusion 59  0860020]}

\makeatletter
\renewcommand\maketitle{
   \begin{center}
     {\huge\sffamily\bfseries\@title\par\vspace{0.3em}}
     {\scshape\large\@author}
   \end{center}
}
\makeatother

\begin{document}

\maketitle

\section*{Abstract}

Pellet ELM triggering is a well established scheme for decreasing the time between two successive ELM crashes below its natural value. Reliable ELM pacing has been demonstrated experimentally in several devices increasing the ELM frequency considerably. However, it was also shown that the frequency cannot be increased arbitrarily due to a so-called lag-time. During this time after a preceding natural or triggered ELM crash, neither a natural ELM crash occurs nor the triggering of an ELM crash by pellet injection is possible. For this article, pellet ELM triggering simulations are advanced beyond previous studies in two ways. Firstly, realistic ExB and diamagnetic background flows are included. And secondly, the pellet is injected at different stages of the pedestal build-up. This allows to recover the lag-time for the first time in simulations and investigate it in detail. A series of non-linear extended MHD simulations is performed to investigate the plasma dynamics resulting from an injection at different time points during the pedestal build-up. The experimentally observed lag-time is qualitatively reproduced. In particular, a sharp transition is observed between the regime where no ELMs can be triggered and the regime where pellet injection causes an ELM crash. Via variations of pellet parameters and injection time, the two regimes are studied and compared in detail revealing pronounced differences in the non-linear dynamics. The toroidal mode spectrum is significantly broader when an ELM crash is triggered enhancing the stochasticity and therefore also the losses of thermal energy along magnetic field lines. In the heat fluxes to the divertor targets, pronounced toroidal asymmetries are observed. In case of high injection velocities leading to deep penetration, also the excitation of core modes like the $2/1$ neoclassical tearing mode is observed.

\section{Introduction}\label{:intro}

\subsection{Motivation} \label{:intro:motivation}
Type-I edge localized modes (ELMs) show an unfavorable scaling towards large machines like ITER both regarding thermal energy losses and the wetted area across which heat loads are distributed at the divertor targets~\cite{Loarte2014} such that ELM control is essential. The application of external resonant magnetic perturbation fields (RMPs) is a promising approach~\cite{Evans2008}, however the applicability has been often found to be restricted to particular “windows” in the edge safety factor q95. Contrary to existing tokamaks, ITER is expected to be in ELMy H-mode also during ramp-up and ramp-down. The RMP operational windows may not allow reliable control in these phases because q95 is evolving during the transient periods. Pellet ELM triggering offers a complementary approach, allowing to increase the ELM frequency and reduce ELM losses~\cite{Lang2004_1, Lang2003, Baylor2013POP}. It is imperative to investigate the lag time after a preceding ELM crash during which ELM triggering by pellets is not possible, since it poses an upper limit for the maximum achievable ELM frequency \cite{Lang_2014}. In the present article, pellet ELM triggering simulations are improved beyond the state of the art, by including realistic plasma background flows and by studying the injection at various time points during pedestal build-up. This is done by injecting pellets during the pedestal build-up of recent simulations of type-I ELM cycles~\cite{Cathey2020}. This way, the transition from a regime where ELM triggering is not possible early in the pedestal build-up into the ELM triggering regime is studied in simulations for the first time. The non-linear features of both regimes are studied and compared in detail. 

The article is structured as follows. Within the present Section~\ref{:intro}, a brief overview is given of the experimental background for pellet ELM triggering (Subsection~\ref{:intro:exp}) and information is given on previous simulations (Subsection~\ref{:intro:sim}). Section~\ref{:setup} explains the simulation set-up used for the present study. The actual simulation results are presented and analyzed in the following Sections. 
First, at constant pellet size and constant injection velocity, the time of injection during the pedestal build-up is varied in Section~\ref{timing_scan} to investigate the transition from no-ELM into the ELM triggering regime. The influence of the injected pellet size onto this transition is further investigated in Section~\ref{size}. Based on these results, the no-ELM and ELM triggering regimes are compared in depth in Section~\ref{compare-regimes} to highlight key differences in the plasma response. The influence of the injection velocity onto the plasma response is analyzed via an additional parameter scan in Section~\ref{velocity}. Finally, conclusions and an outlook are provided in Section~\ref{conclusions}. References and Acknowledgements follow at the very end of the article.

\subsection{Experimental background}\label{:intro:exp}

At ASDEX Upgrade in the divertor DIV IIb~\cite{Herrmann2003} configuration and a vessel wall surface covered with about half of carbon and half of tungsten (AUG-C), first pioneering experiments on ELM pacing and mitigation by pellet injection were performed \cite{Lang2004_1}. 
For cases where the pellet injection rate $f_p > 1.5 \times f^0_\mathrm{ELM}$ with $f^0_\mathrm{ELM}$ the natural ELM frequency, full ELM frequency control with $f_\mathrm{ELM} = f_p$ was achieved \cite{Lang2004_1, Lang2007}. Furthermore, albeit only with $f_\mathrm{ELM}$ in the range 50 – 110 Hz, $dW_\mathrm{ELM}/W_0 \sim 1/f_\mathrm{ELM}$ 
was found, with $dW_\mathrm{ELM}$ the ELM induced plasma energy loss and $W_0$ the pre-ELM plasma energy \cite{Herrmann2003}. 
However, at high pellet injection frequencies, unwanted increases of the plasma density were observed \cite{Lang_2014}. For the investigated experimental conditions, every pellet injected during an H-mode phase triggered an ELM within less than 0.3 ms after reaching the separatrix \cite{Lang2006}. Main findings of AUG-C were confirmed at other machines like DIII-D \cite{Baylor2013PRL} and JET \cite{Romanelli2009}. 
In DIII-D, small pellets (1.3 mm cylindrical pellet) triggered small ELMs within 0.1 ms of the pellet entering the plasma \cite{Baylor2013POP}. 
The ELM event is found within 1 cm of the pellet crossing the separatrix, while slightly shallower than what is observed in AUG-C, 3 cm \cite{Kocsis2007}. Spontaneous and triggered ELMs were shown to have very similar properties \cite{Lang2008}.
Motivated by these findings, ELM control by pellets was re-visited in the tungsten ASDEX Upgrade configuration (AUG-W) \cite{Lang_2014}. A dedicated analysis of a specific plasma scenario showed that successful ELM triggering entails a lag time. During such lag time, injected pellets fail to trigger an ELM crash. This poses an upper limit on the achievable ELM pacing frequency. The experimental observation of the lag time in AUG-W did not show clear correlations to the imposed magnitude of the pellet perturbation, i.e. different pellet sizes or velocities. However, under different plasma conditions, cases were found with pellets failing to trigger ELMs although the pedestal had almost fully recovered from the previous ELM crash. On the other hand, sometimes another ELM was initiated very shortly after an ELM with yet the energy drop still present. This indicates that pedestal stability is not monotonically decreasing over the ELM cycle. The observation of pellet-triggered ELMs without a pronounced pedestal was repeated in attempts to achieve ELM control at the L-H transition by means of pellet pacing both in the AUG-W and JET all-metal-wall tokamaks~\cite{Lang2015}.
As well as the study of lag-time, the pellet-induced plasma responses are also important to be investigated since it may provide the physics understanding of the observation. The magnetic perturbations spectra and the toroidal mode number have been analyzed for JET plasma using a wavelet analysis. It is found that the magnetic perturbations induced by pellets have distinct frequencies and toroidal mode numbers \cite{Poli2010}. The experimental study of pellet-induced magnetic perturbations in ASDEX Upgrade reveals that the magnitude of the pellet-driven perturbation increases monotonically with pellet penetration, and showed an exponential decay after pellet burn-out \cite{Szepesi2009}. The analysis of the ELM onset using magnetic pick-up coil signals shows that a pellet has to reach a certain magnetic surface of the plasma, and the most probable location is in the middle of the pedestal where the pressure gradient is very large \cite{Kocsis2007}.

\subsection{Previous simulations}\label{:intro:sim}
Theoretical and numerical approaches to understand the non-linear MHD physics in response to a pellet injection is a high priority research topic. There are extensively elaborated extended non-linear MHD codes world wide such as NIMROD~\cite{Sovinec2004}, M3D-C1~\cite{Ferraro2010},  BOUT++~\cite{Dudson2009}, JOREK~\cite{Huysmans2007} in order to understand the physics in realistic geometry.
The physics of edge localized modes (ELMs) and ELM control by RMPs, QH-mode (Quiescent H-mode), vertical magnetic kicks, and pellets has been investigated already via non-linear simulations in many ways using the JOREK code~\cite{Huysmans2007, Becoulet2014, Hoelzl2020A, Orain2013, Artola2018, LiuF2015, Futatani2014}. 
Very recently, type-I ELM cycles and the triggering mechanism responsible for the violent onset of the ELM crash were studied for the first time \cite{Cathey2020}. These ELM cycle simulations form the basis for investigating pellet injection at various times during the inter-ELM phase in the present article. The injection of pellets into ASDEX Upgrade for ELM triggering had not been simulated before. However, the injection of deuterium shattered pellets for disruption mitigation was already studied using similar physics models like they are applied in the present article \cite{Hoelzl2020A}. 
The injection of pellets for ELM control has also been studied with M3D-C1; for hydrogenic pellets~\cite{Diem2019} and for Lithium Granule Injection in linear simulations~\cite{Fil2017}. Further work in particular on impurity pellet injection exists, but is aiming at disruption mitigation.
First simulations using JOREK for the triggering of an ELM crash by the injection of a frozen deuterium pellet modelled as a localized, static density perturbation were shown in Ref.~\cite{Huysmans2009}. Afterwards, results based on a spatio-temporally varying pellet ablation model were shown including experimental comparisons to DIII-D~\cite{Futatani2014} and JET~\cite{Futatani2019}. The pellet size requirement for ELM triggering in the stable plasma was found to be $\sim 70$\% of the minimum pedestal pressure which causes spontaneous ELM. The key parameter of ELM control by pellet injection is the three-dimensionally localized pressure perturbation at the plasma edge, and the physics understanding is continued to be revealed by theory and numerical simulations ~\cite{Huysmans2009, Futatani2014}. 
In experiments, pellets can be only launched at certain time slots, therefore the timing of pellets reaching the plasma is difficult to choose. In the study of pellet-triggered ELMs in JET, the pellets were injected into the unstable plasma slightly before the spontaneous ELM event. The JOREK simulations showed good agreement with the experimental observations of the heat flux reaching the plasma facing components (PFC), $\sim 60$ $\mathrm{MW/m^2}$ \cite{Frigione2015}. The magnitude of the peak of the heat flux is similar between the spontaneous ELM and the pellet-triggered ELM, also consistent with the experiment. Furthermore, a toroidally asymmetric heat deposition onto the divertor targets related to pellet-triggered ELMs has been observed in the previous simulations for DIII-D and JET \cite{Futatani2014, Futatani2019} consistent with experimental observations \cite{Wenninger2011}. 

This article extends previous work in several ways. The first pellet ELM triggering simulations are presented for ASDEX Upgrade, realistic ExB and diamagnetic background flows are included for the first time, and the injection during different phases of pedestal build-up is studied for the first time. This allows detailed insights into the experimentally observed lag-time. A direct comparison between simulations of spontaneous and pellet-triggered ELMs is not part of this work, but is studied separately.

\section{Simulation setup}\label{:setup}

\begin{figure}
\centering
  \includegraphics[width=0.7\textwidth]{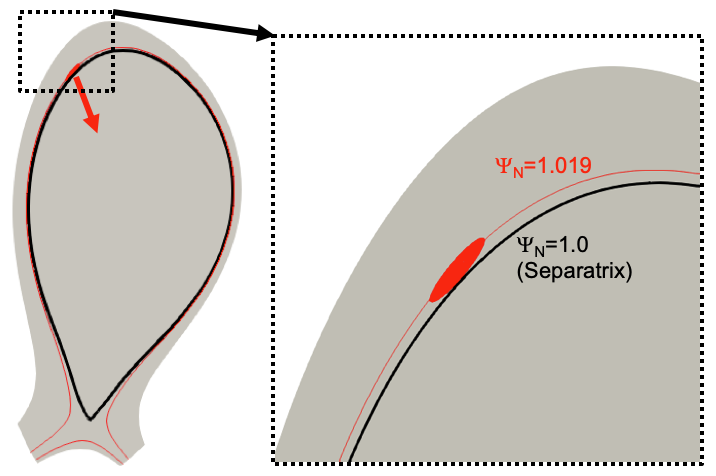}
\caption{The pellet trajectory is started on the realistic experimental trajectory just outside the separatrix (to save computational time). The red arrow indicates the pellet trajectory into the plasma.}
\label{fig:pelletlocation}
\end{figure}

Pellet ELM pacing experiments in the ASDEX Upgrade tokamak~\cite{Meyer2019} are performed using a system which injects pellets from top of the High Field Side (HFS) as shown in Ref.~\cite{Lang_2014}. Pellets are prepared in the cryogenic system and transported to the HFS via a 17 meter long guiding tube. The technical capabilities of the pellet injector at  ASDEX Upgrade are the following: an injection frequency up to 70~Hz is possible, the pellet size can range from $1.5 \times 10^{20}$ to $3.7 \times 10^{20}$ [particles/pellet], and the injection velocity can range between 240 and 1040 m/s (dependent on the pellet size). 
In the simulations, the pellet size is the number of atoms contained in a pellet. For example, `$0.8 \times 10^{20}$D' is a pellet which contains $0.8 \times 10^{20}$ deuterium atoms.
In the simulations, the initial pellet location is R= 1.365 [m], Z=0.6737 [m], where the normalized poloidal magnetic flux is $\Psi_N=1.019$. 
It corresponds to 1.8 cm outside the separatrix as shown in Fig~\ref{fig:pelletlocation}.

Experimentally, it is considered that about 50 \% of the pellet particles are lost in the 17 m long pellet guide~\cite{Lang2003} such that our base simulations carried out with pellets containing $0.8 \times 10^{20}$ Deuterium atoms correspond approximately to the smallest pellet size possible experimentally. In addition, simulations with $0.4 \times 10^{20}$ and $1.5 \times 10^{20}$ atoms are carried out to investigate size dependencies in section~\ref{size}.

Simulations are carried out with the non-linear MHD code JOREK~\cite{Huysmans2007} based on a fully implicit time stepping, and a spatial discretization with 2D Bezier elements in the poloidal plane combined with a toroidal Fourier expansion~\cite{Czarny2008}. The extended physics model including ExB and diamagnetic background flows described in Ref.~\cite{Orain2013} is used. Mach-1 boundary conditions are applied at the divertor targets to model the plasma sheath at the divertor targets. The injection is performed at different time points during the ELM cycle simulations which are described in Ref.~\cite{Cathey2020}. Details of the ablation and pellet model are described in Refs.~\cite{Futatani2014, Futatani2019}.

\begin{figure}
\centering
  \includegraphics[width=0.8\textwidth,height=0.6\textwidth]{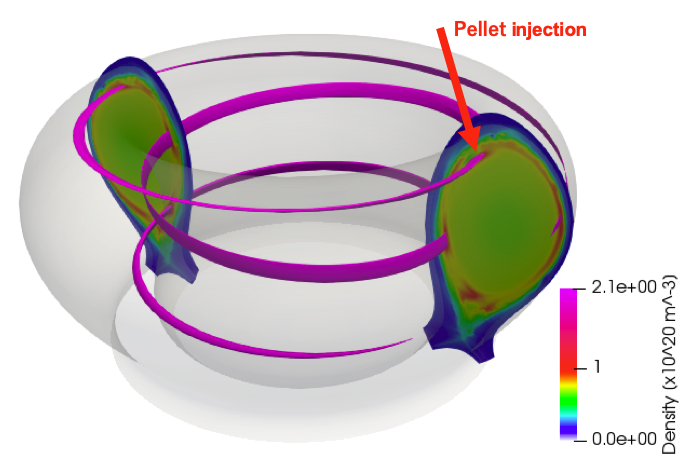} 
\caption{The pellet cloud (density perturbation) which is defined by $1.3 \times 10^{20}~\mathrm{m^{-3}}$ is plotted in pink band at the time of the maximum ablation rate $t=12.274$ ms ($t_\mathrm{injection}=12$~ms). The pellet cloud which originates the pellet location propagates along the magnetic field line. }
\label{fig:08D_12ms_24560}
\end{figure}

The mechanism of pellet ELM triggering is illustrated based on the simulation with the pellet containing $0.8 \times 10^{20}$ deuterium atoms injected at 12 ms during the pedestal build-up which is explained in detail later on. Figure~\ref{fig:08D_12ms_24560} shows the high density pellet cloud (pink band). The contour is plotted at $1.3 \times 10^{20}~\mathrm{m^{-3}}$ at the time of the maximum ablation rate $t=12.274$ ms which is one of the time slice of the inter-ELM, described in Section~\ref{timing_scan}.
The pellet ablation is adiabatic and, as it proceeds, the localized high density region created by the ablation of the pellet expands along field lines with the local sound speed. The pellet cloud is heated by the electrons along the magnetic field line with the parallel thermal diffusion which is much faster than the pellet cloud expansion. 
The resulting local high pressure perturbation is responsible for the ELM onset. 

In this manuscript, the pellet conditions such as pellet size and injection speed have been varied to analyze the dependence. To refer more easily to the different configurations, the pellet conditions are named as summarized in Table~\ref{table:names_size} and Table~\ref{table:names_speed}.

\begin{table}[htbp]
\caption{The names of the pellet sizes used in this work. }
\label{table:names_size}
\centering
\begin{tabular}{|c|c|}
\hline 
Name for size & Particle content in a pellet \\
\hline
Small & $0.4 \times 10^{20}$\\ 
\hline
Medium & $0.8 \times 10^{20}$\\ 
\hline
Large & $1.5 \times 10^{20}$\\ 
\hline
\end{tabular}
\end{table}

\begin{table}[htbp]
\caption{The names of the pellet velocities used in this work. }
\label{table:names_speed}
\centering
\begin{tabular}{|c|c|}
\hline 
Name for velocity & Pellet injection velocity \\
\hline
Very slow &  240 m/s \\ 
\hline
Slow &  300 m/s\\ 
\hline
Medium &  560 m/s\\ 
\hline
Fast &  800 m/s\\ 
\hline
\end{tabular}
\end{table}
\section{Pellet injections at different times during pedestal build-up}\label{timing_scan}

\begin{figure}
\centering
  \includegraphics[width=0.7\textwidth]{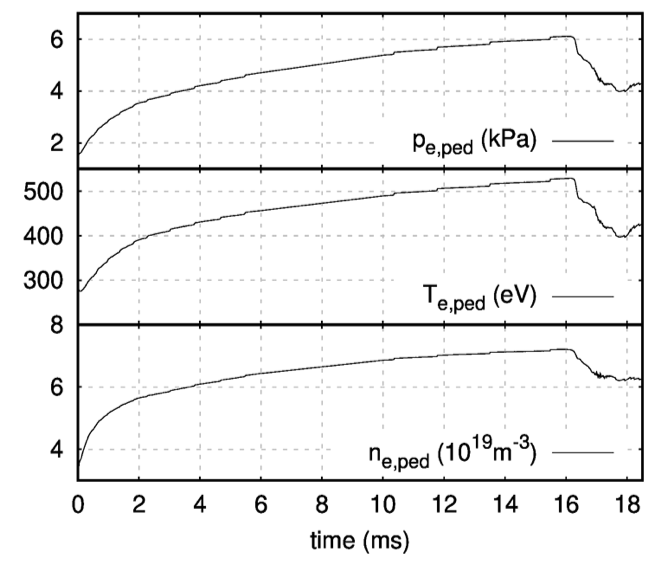}
\caption{The time evolution of the pressure, the temperature and the density at the pedestal top of the base natural ELM case taken from the series of simulations described in Ref.~\cite{Cathey2020}. Pellet injection simulations are modelled at different times during the build-up phase. Losses from the natural ELM can be seen to start at about 16.1 ms. }
\label{fig:pedestalbuild}
\end{figure}

\begin{figure}
\centering
\includegraphics[width=0.7\textwidth]{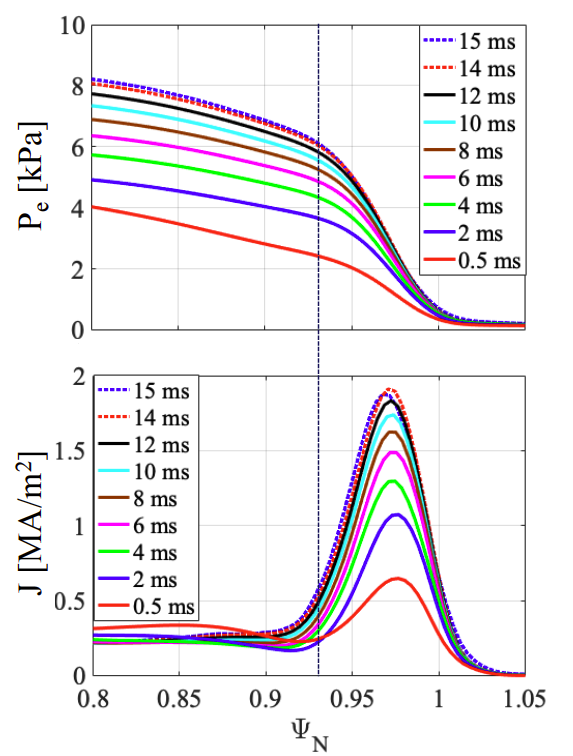} 
\caption{Profiles of toroidal averaged electron pressure ($p_e$) and current density ($J$) in the pedestal region for the injection times 0.5, 2, 4, 6, 8, 10, 12, 14 and 15 ms. The pedestal top $\Psi_N \sim 0.932$ is indicated by a dashed line.}
\label{fig:pedestal_profiles_time_full}
\end{figure}

In this entire section, we focus only on simulations with a pellet size corresponding to $0.8 \times 10^{20}$D atoms injected at 560 m/s while considering different injection times. Figure~\ref{fig:pedestalbuild} shows the time evolution of pressure, temperature and density at the pedestal top during the inter-ELM period in the JOREK ELM cycle simulation used as basis for the present study. The post-ELM profiles build up until they reach the MHD stability limit and a natural ELM crash eventually occurs at about 16 ms causing a significant loss of particles and thermal energy~\cite{Cathey2020}.
Figure~\ref{fig:pedestal_profiles_time_full} shows the profiles of toroidally averaged pedestal pressure and the current density for the time slices 0.5 ms, 2 ms, 4 ms, 6 ms, 8 ms, 10 ms, 12 ms, 14 ms and 15 ms. Table~\ref{table:pedestal_evolve} shows the pellet injection timings and corresponding pedestal parameters. Here, the pedestal top is considered at $\Psi_N \sim 0.932$ and the peak of the current density profiles at the outer midplane are given at $\Psi_N \sim 0.973$.
\begin{table}[htbp]
\caption{The pellet injection timings and corresponding pedestal parameters (the total pressure ($p=p_i+p_e$) at the pedestal top $\Psi_N \sim 0.932$ and the current density at $\Psi_N \sim 0.973$).}
\label{table:pedestal_evolve}
\centering
\begin{tabular}{|c|c|c|c|}
\hline 
inj. time & total pressure $p$ & current density $J$ & max.$ |\nabla p| $ \\
\hline
 0.5 ms  & 4.8  kPa & 0.65 $\mathrm{MA/m^2}$ & 260.6 kPa/m\\ 
 2 ms    & 7.4  kPa & 1.07 $\mathrm{MA/m^2}$ & 424.1 kPa/m\\
 4 ms    & 8.6  kPa & 1.30 $\mathrm{MA/m^2}$ & 499.3 kPa/m\\
 6 ms    & 9.6  kPa & 1.49 $\mathrm{MA/m^2}$ & 575.7 kPa/m\\
 8 ms    & 10.6  kPa & 1.63 $\mathrm{MA/m^2}$ & 624.5 kPa/m\\
 10 ms   & 11.2  kPa & 1.74 $\mathrm{MA/m^2}$ & 662.3 kPa/m\\
 12 ms   & 11.6  kPa & 1.83 $\mathrm{MA/m^2}$ & 698.2 kPa/m\\
 14 ms   & 12.0  kPa & 1.91 $\mathrm{MA/m^2}$ & 727.5 kPa/m\\
 15 ms   & 12.2  kPa & 1.88 $\mathrm{MA/m^2}$ & 711.0 kPa/m\\
 \hline
\end{tabular}
\end{table}

\begin{figure}
\centering
\begin{tabular}{c c}
  \includegraphics[width=0.5\textwidth]{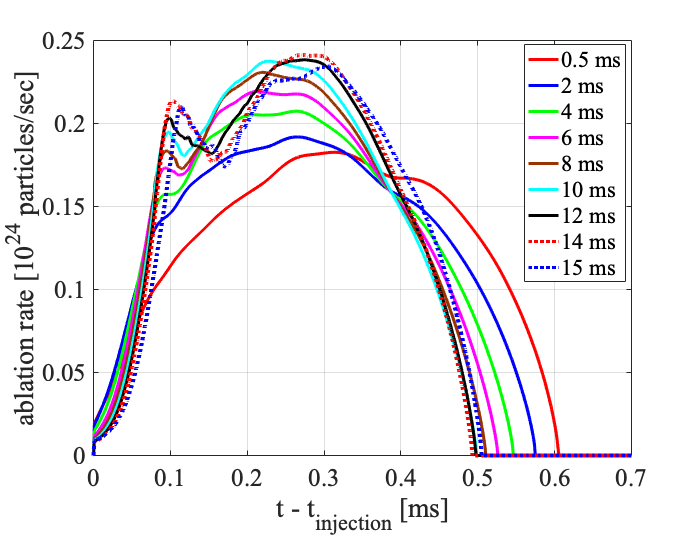} &
  \includegraphics[width=0.5\textwidth]{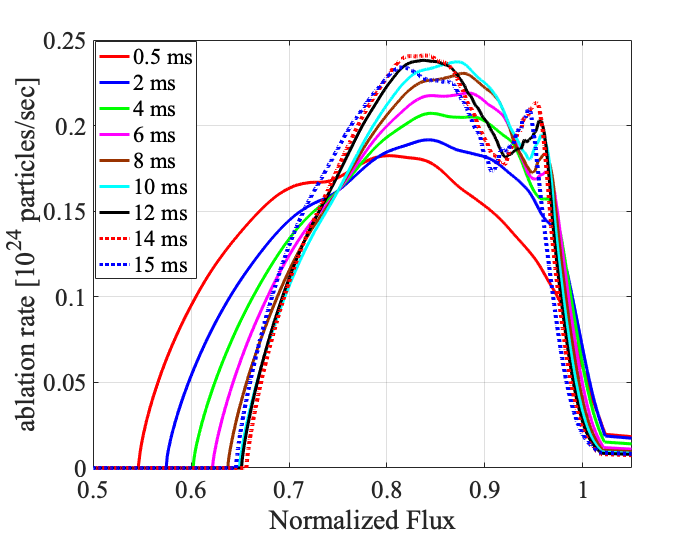} 
\end{tabular}
\caption{(a) The time evolution of the pellet ablation rate for different pellet injection timings versus $t - t_\mathrm{injection}$. The pellet size is $0.8 \times 10^{20}$D and the injection velocity is 560 m/s. (b) The pellet ablation rate versus normalized poloidal magnetic flux. The pellet locations at the ELM onset (of cases which ELMs are triggered) are $\Psi_{N,pellet} =0.956$ ($t_\mathrm{injection}$ = 12 ms), $\Psi_{N,pellet} =0.952$ ($t_\mathrm{injection}$ = 14 ms), and $\Psi_{N,pellet} =0.945$ ($t_\mathrm{injection}$ = 15 ms) as summarized in Table~\ref{table:ELMonset}.}
\label{fig:ablataionrate_08D}
\end{figure}

Pellet injections are simulated at different times during build-up which correspond to evolving MHD stability conditions (0.5 ms, 2 ms, 4 ms, 6 ms, 8 ms, 10 ms, 12 ms, 14 ms and 15 ms). 
Table~\ref{table:08D_summary} summarizes the information of the maximum ablation rate and the full ablation rate for the dependence of pellet injection timings. 
\begin{table}[htbp]
\caption{List of the simulation cases which are performed for this section with information regarding time of the maximum ablation rate and the full ablation.}
\label{table:08D_summary}
\begin{tabular}{|c|c|c|c|c|}
\hline 
inj. time & Time at  max. abln. & $\Psi_N$ at max. abln. & Time at full abln. & $\Psi_N$ at full abln. \\
\hline
 0.5 ms & 0.8088 ms & 0.80 & 1.1013 ms & 0.5456 \\ 
 2 ms & 2.2635 ms & 0.842 & 2.573 ms & 0.574 \\
 4 ms & 4.2635 ms & 0.845 & 4.546 ms & 0.6018 \\
 6 ms & 6.2136 ms & 0.8835 & 6.527 ms & 0.6216 \\
 8 ms & 8.2164 ms & 0.8781 & 8.5068 ms & 0.6374 \\
 10 ms & 10.226 ms & 0.872 & 10.496 ms & 0.6493 \\
 12 ms & 12.274 ms & 0.8376 & 12.4967 ms & 0.6513 \\
 14 ms & 14.268 ms & 0.8384 & 14.488 ms & 0.6562 \\
 15 ms & 15.3045 ms & 0.813 & 15.5035 ms & 0.6456\\
 \hline
\end{tabular}
\end{table}
Figure~\ref{fig:ablataionrate_08D}(a) and (b) show the ablation rates versus time ($t - t_\mathrm{injection}$) and versus normalized poloidal magnetic flux, respectively for the different injection times. The ablation process for this pellet size completes within 0.5-0.6 ms depending on the chosen injection time. The pellet reaches the separatrix in 0.033 ms which is very fast compared to the time-scale of MHD activities and pellet ablation physics which we are looking at. When the pellet crosses the separatrix, it starts ablating according to the local density and temperature. The pellet injection timing gives different pellet ablation rate, in terms of the maximum ablation rate and the pellet penetration depth due to the different plasma parameters at those times. Specifically, pellet ablation for early injections in the cycle (0.5 ms, 2 ms, 4 ms, 6 ms and 8 ms) takes longer than for late-injection cases since the temperature is lower. As a consequence, early injection cases show deeper pellet penetration into the plasma, reaching ${\Psi_N < 0.65}$. In case of later injection, the ablation rate starts to drop around ${\Psi_N\approx0.95}$ due to the collapse (relaxation) of the pedestal structure (the ELM crash). Still, the penetration depth in case of late injection remains shallower due to the higher ablation rate in the hotter plasma.

The pellet location at the time of the ELM onset in cases where an ELM is triggered is approximately $\Psi_\mathrm{N,pellet}=0.956$, $0.952$ and $0.945$ for injections at 12, 14, and 15 ms respectively, as summarized in Table~\ref{table:ELMonset}. Thus, only a weak dependency on the injection time is observed. The ELM onset is triggered when the pellet is close to the top of the pedestal steep gradient region roughly in line with experimental observations, which suggest a triggering in the middle of the pedestal in the steep pressure gradient region~\cite{Kocsis2007}.

\begin{figure}
\centering
\begin{tabular}{c c}
  \includegraphics[width=0.48\textwidth]{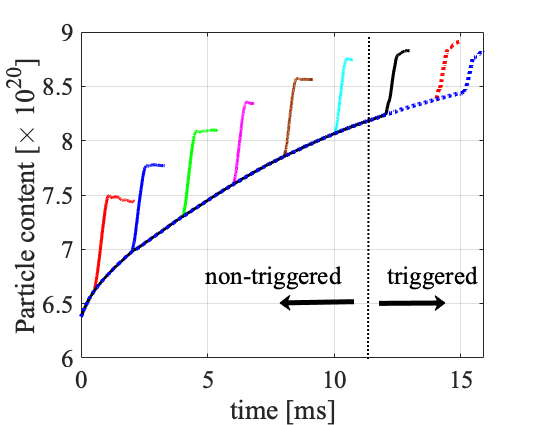} &
  \includegraphics[width=0.48\textwidth]{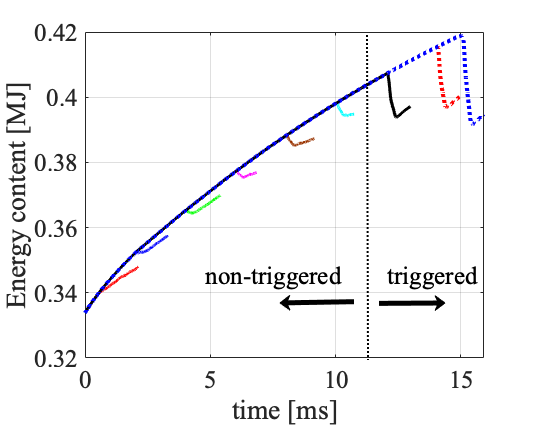} 
\end{tabular}
\caption{(a) The particle and (b) the energy content inside of the separatrix versus time for different pellet injection timings. The pellet size is $0.8 \times 10^{20}$D and the injection velocity is 560 m/s.}
\label{fig:contents_lagtime08}
\end{figure}

\begin{figure}
\centering
\begin{tabular}{c c}
  \includegraphics[width=0.5\textwidth]{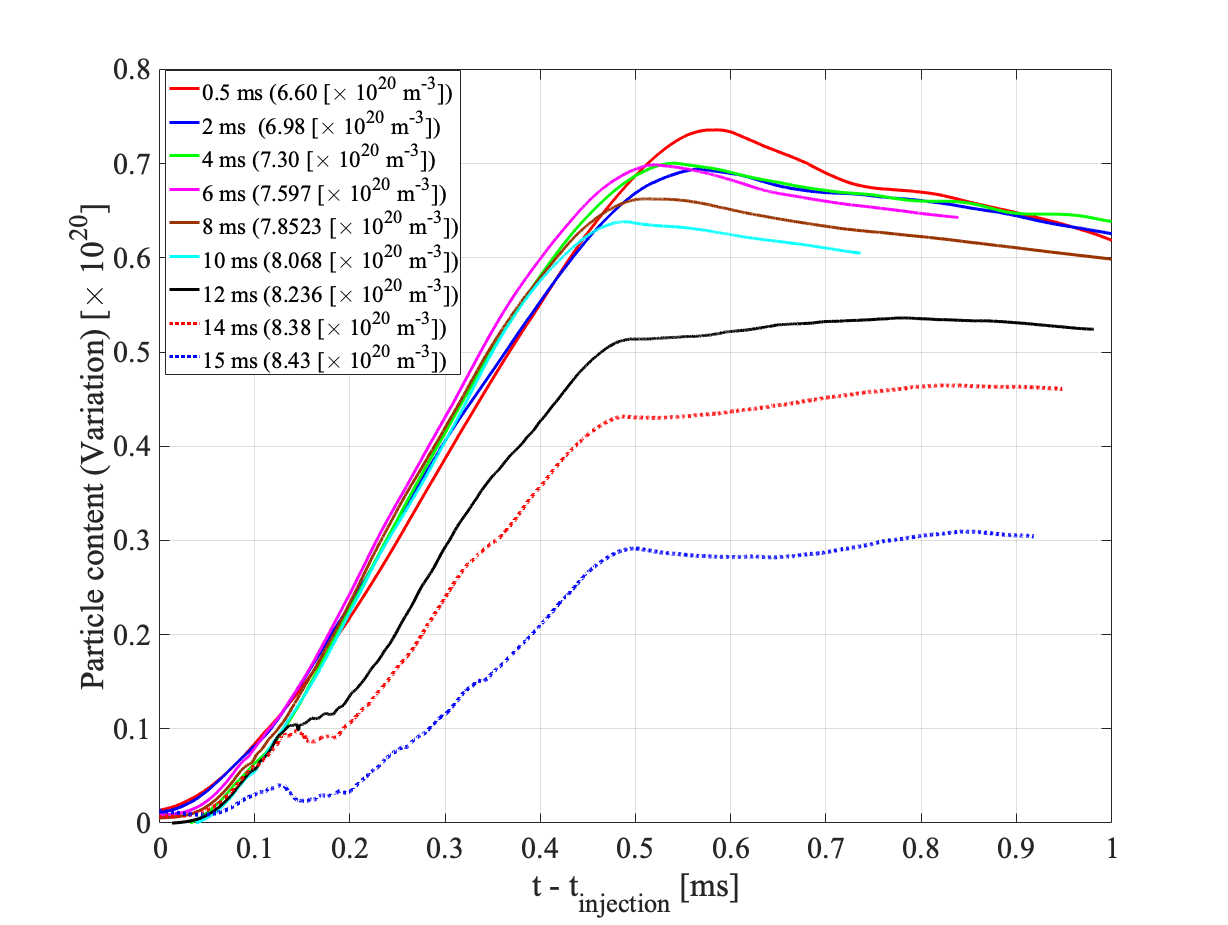} &
  \includegraphics[width=0.5\textwidth]{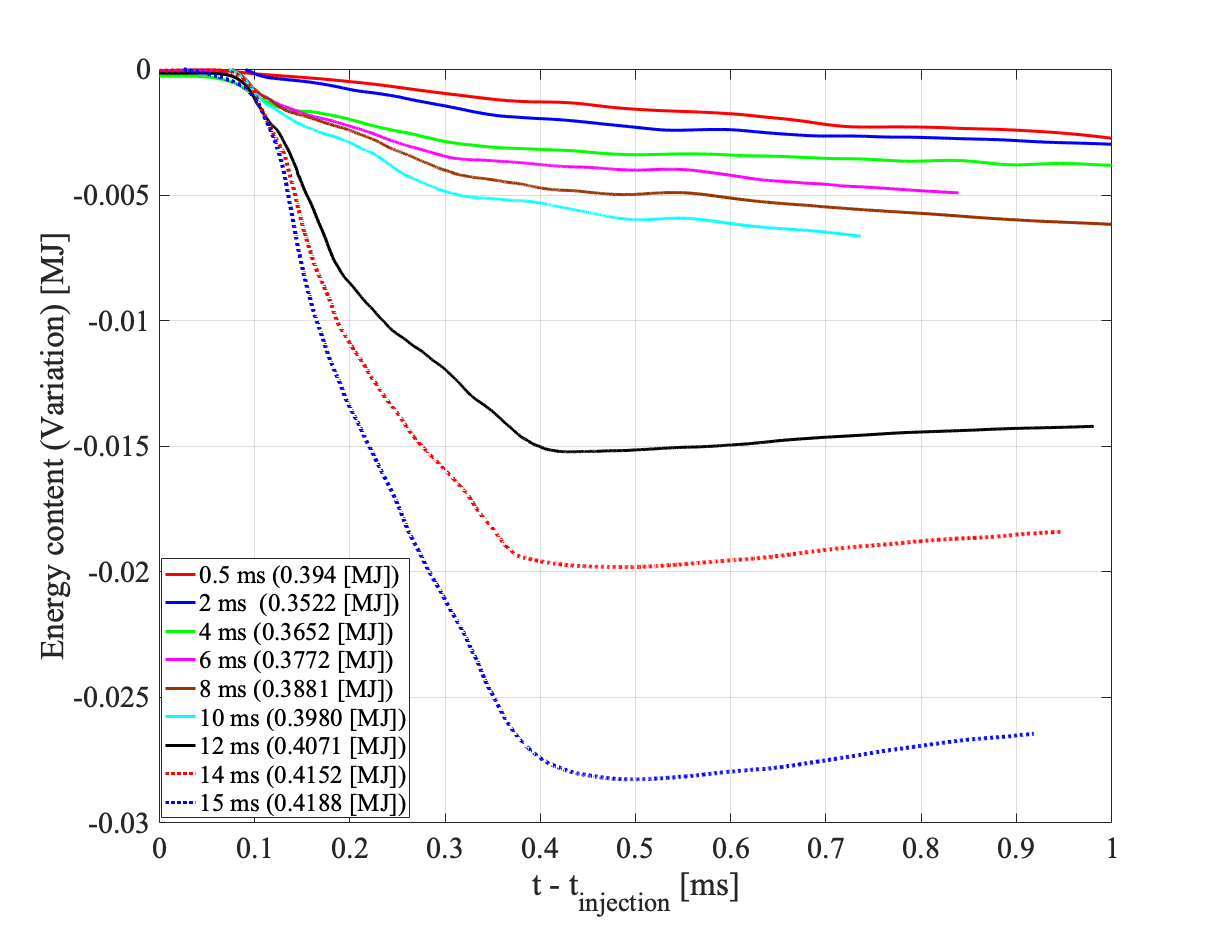} 
\end{tabular}
\caption{(a) The particle and (b) the energy content inside the separatrix versus time for different pellet injection times. The x-axis is shifted with respect to the injection time. For the y-axis, the difference between the injection case and an equivalent axisymmetric simulation without injection is plotted.}
\label{fig:normcontents_lagtime08_shifted}
\end{figure}

Figure~\ref{fig:contents_lagtime08} shows the particle and the energy content inside of the separatrix versus time for different pellet injection times. 
The time evolution of the particle content clearly indicates that the pellets deliver particles into the plasma. Contrary to early pellet injections, in the cases of late injections ($\geq$ 12 ms), the increase of the particle content inside the separatrix is significantly below the pellet content ($0.8 \times 10^{20}$) since the pellet-triggered ELM expels particles from the plasma. 
For the same reason, a strong drop of the plasma thermal energy content is observed when the pellet is injected at 12 ms or later.
In order to compare the variation of the particle and the energy content inside the separatrix, the evolutions are shifted with respect to the values at the times of pellet injection as shown in Fig.~\ref{fig:normcontents_lagtime08_shifted}. 
There is a sharp transition in the thermal energy drop between the pellet injection timings of 10 ms and 12 ms, indicating the transition from no-ELM to pellet ELM triggering. 

\begin{figure}
\centering
\includegraphics[width=0.6\textwidth]{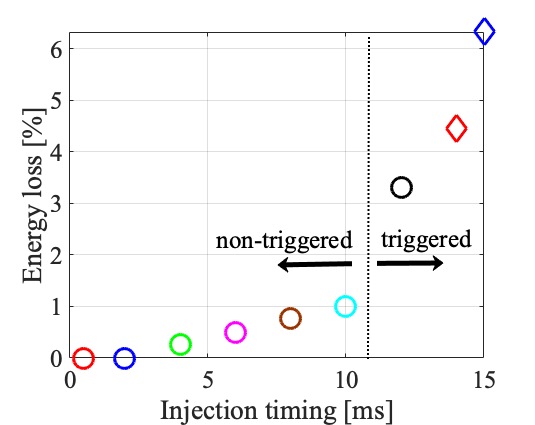} 
\caption{The thermal energy loss in percentage for different pellet injection times. The pellet size is $0.8 \times 10^{20}$D and the injection velocity is 560 m/s.}
\label{fig:energyloss_08D}
\end{figure}

Figure~\ref{fig:energyloss_08D} shows the relative loss of plasma thermal energy for different pellet injection times. The relative energy loss is measured as the difference between the maximum value before the crash and the minimum value before the thermal energy starts to increase again and is then normalized by the total plasma thermal energy before the crash to obtain the relative value. Pellet injections at very early timings, 0.5 ms and 2 ms do not show energy losses at all according to this definition, i.e. there is no minimum in the thermal energy content. However, for these cases the injections still cause reductions in comparison to the case without injection as seen in Figure~\ref{fig:contents_lagtime08}. The injection timings of 4 ms, 6 ms, 8 ms, and 10 ms show energy losses of $\leq$ 1 \%. There is a sharp transition in the thermal energy loss between cases where no ELM is triggered (0.5 - 10 ms) and cases where an ELM is triggered (12 ms - 15 ms). 
The injection times should not be compared one to one to the experiment, since the pedestal build-up might not be identical. Instead, we analyze the pedestal parameters corresponding to the transition between the no-ELM triggering and ELM triggering regimes in Figure~\ref{fig:pedestal_profiles_time_full}.

\begin{figure}
\centering
 \includegraphics[width=0.5\textwidth]{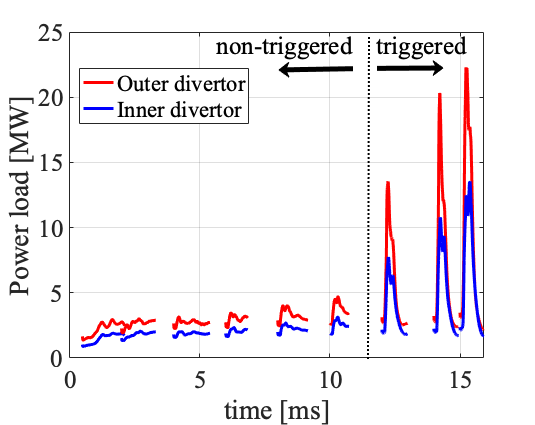} 
\caption{The time evolution of the power load onto the inner and the outer divertor targets which is caused by $0.8 \times 10^{20} D$ pellet injections for the various injection times.}
\label{fig:powers_08D}
\end{figure}

Figure~\ref{fig:powers_08D} shows the power load onto the inner and the outer divertor targets which is caused by the pellet injections. Most of the power goes to the divertor targets. There is a sharp transition in the peak of the integrated power load onto the divertor targets between cases where no ELM is triggered ($\leq$ 10 ms) and cases where ELMs are triggered ($\geq$ 12 ms). The peak of the power load onto the outer divertor target in no-ELM triggered cases ($\leq$ 10 ms) is $\leq$ $5~\mathrm{MW}$. The peak of the power load onto the outer divertor target where ELMs are triggered ($\geq$ 12 ms) reaches $\geq$ 10 MW, especially late injection cases ($\geq$ 14 ms) shows $\geq$ 20 MW power load onto the outer divertor target. A strong increase of the power load is observed to last roughly 0.4 ms due to the pellet-triggered ELM. 
The power load onto the outer divertor target is twice larger than the one onto the inner target. The observation is universal for the simulations performed in this study.
The distribution of the heat between the targets should not be compared directly to the experiment, since the SOL model used in the simulations is very simplified and remains to be enhanced in future studies. 

\begin{figure}
\centering
  \includegraphics[width=0.8\textwidth]{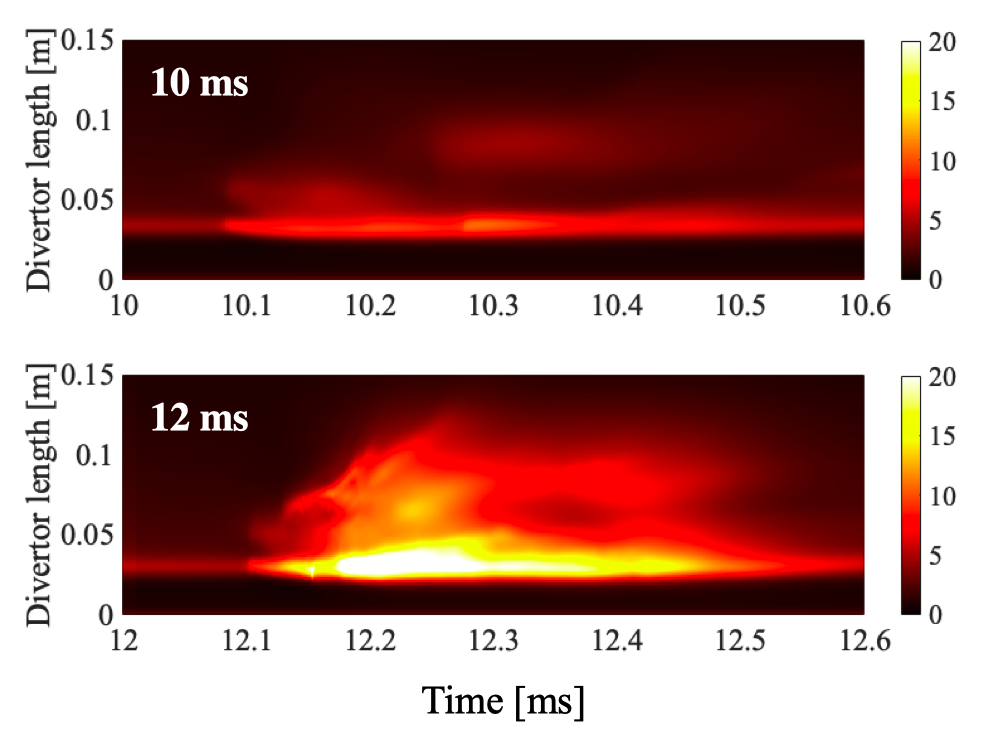} 
\caption{The time evolution of the heat flux onto the outer divertor targets which is caused by $0.8 \times 10^{20} D$ pellet injection with the injection velocity of 560 m/s. Two cases are compared; no-ELM triggering case (10 ms) and pellet-triggered ELM case (12 ms).}
\label{fig:heatflux_lagtime_08D}
\end{figure}

The pellet is injected at the toroidal angle $\varphi = 0$. 
The heat flux profile onto the outer divertor target at $\varphi = 0$ versus time is shown in Fig.~\ref{fig:heatflux_lagtime_08D} for the pellet injection times of 10 ms and 12 ms. 
It is clearly visible that the case without ELM triggering does not show a prominent increase of heat flux onto the divertor. On the other hand, the case of pellet injection at 12~ms shows a strong increase of the heat flux $\sim 20~\mathrm{MW/m^{2}}$ at the strike point for 12.1~ms - 12.5~ms. 
There is some correlation with the ablation rate since the ablation rate directly determines the pressure perturbation and mode excitation even in the no-triggering cases. These modes lead to some edge stochastization close to the separatrix and immediately cause losses. There are three different deposition locations which might be related to magnetic tangles forming in the stochastic field.

\begin{figure}
\centering
\begin{tabular}{c c}
  \includegraphics[width=0.5\textwidth]{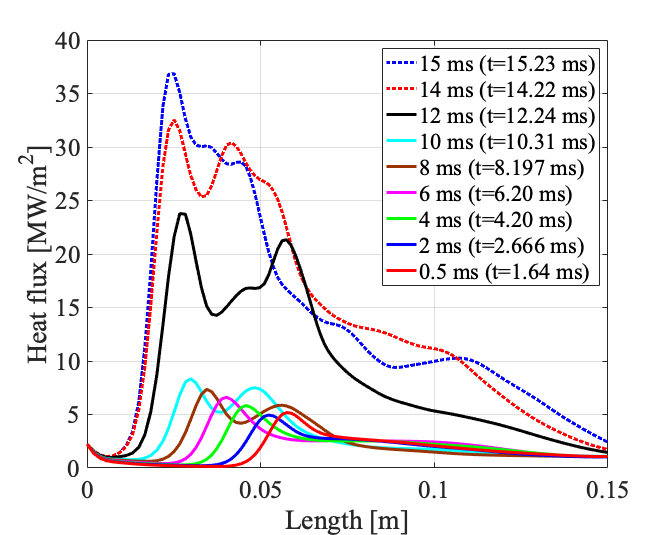} &
  \includegraphics[width=0.5\textwidth]{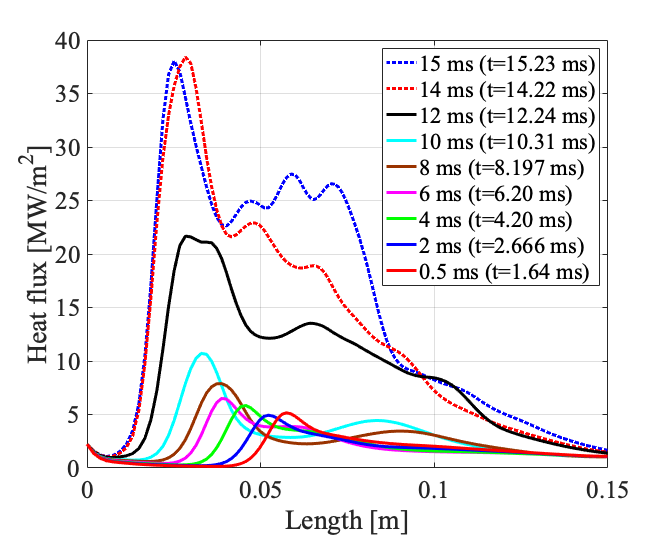} 
\end{tabular}
\caption{Heat flux profiles of the outer divertor target at maximum power load onto the outer divertor target for the various cases with different injection times (see Fig.~\ref{fig:powers_08D}). The left panel shows the heat flux profiles at a toroidal angle of 180 degrees ($\varphi = 180^\circ$) and the right panel shows the ones at the toroidal angle of 0 degrees ($\varphi = 0^\circ$), i.e., at the toroidal angle of pellet injection.}
\label{fig:heatflux_maxload_08D}
\end{figure}

Figure~\ref{fig:heatflux_maxload_08D} shows the heat flux profile at the maximum power load onto the outer divertor target (see Fig.~\ref{fig:powers_08D}). The early injection cases (0.5 - 10 ms) which do not trigger an ELM show a peak heat flux of ${\leq 10~\mathrm{MW/m^{2}}}$. The cases which do trigger ELMs (${t_\mathrm{inj.}\geq12~\mathrm{ms}}$) show a peak heat flux of ${\geq 20~\mathrm{MW/m^{2}}}$. Peak heat fluxes for pellet injection at 14 ms and 15 ms are ${\geq 30~\mathrm{MW/m^{2}}}$. The heat flux profiles of these pellet-triggered ELMs display toroidally asymmetric characteristics which are described in the following paragraph. The wetted area as well as many other non-linear features of the simulated pellet-triggered ELMs in comparison to spontaneous ELMs are under investigation.

\begin{figure}
\centering
  \includegraphics[width=0.8\textwidth]{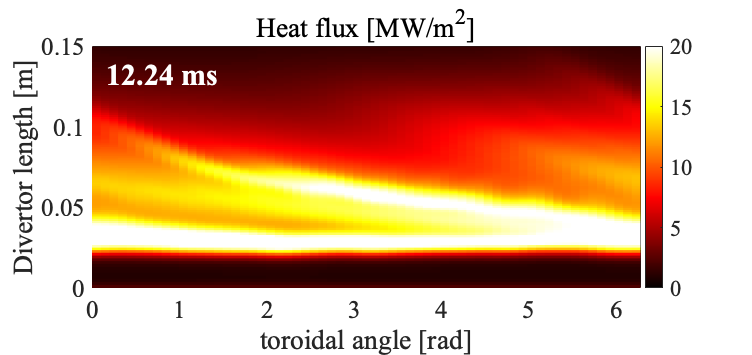} 
\caption{The heat flux profile along the toroidal direction in the case with injection at 12 ms, which triggers an ELM. The outer divertor target is shown at $t=12.24$ ms (time of maximum heat flux). A strongly asymmetric structure is observed with a clear strike-line splitting.}
\label{fig:heatflux_08D_12ms_toroidalangle}
\end{figure}

Toroidally asymmetric features of the pellet-triggered ELM and a sub-structure in the heat deposition with several peaks are observed, and are qualitatively similar to previous simulations for JET~\cite{Futatani2019}.
The heat flux profile versus the toroidal angle in the case with $0.8 \times 10^{20}$D pellet injection at ${t_\mathrm{inj.}=12~\mathrm{ms}}$ is shown in Fig.~\ref{fig:heatflux_08D_12ms_toroidalangle}. 
The time slice of ${t=12.24~\mathrm{ms}}$ corresponding to the maximum power load onto the plasma facing components is plotted. A strongly asymmetric profile of the heat flux profile is observed. 
The observation of an ${n=1}$ toroidal structure is universal in all cases of pellet-triggered ELMs in this study. 
The toroidally asymmetric features observed in the heat flux deposition profiles are caused by the 3D helical perturbation induced by the pellet and its ablation dynamics.

In this section, the plasma response to pellet injection at different times during pedestal build-up was analysed. 
The simulations in this section were all carried out with pellets containing $0.8 \times 10^{20}$ deuterium atoms and an injection velocity of 560 m/s in order to investigate the change of the MHD stability during the pedestal build-up. 
The simulations clearly show a sharp transition between early injection with moderate losses and divertor heat fluxes (lag-time) and later injections with an explosive onset of MHD instabilities causing strong losses and large divertor heat fluxes.  The plasma parameters in terms of the pedestal pressure of $p_\mathrm{ped} = 11.2$ and 11.6 kPa which corresponds to the injection timing of 10 ms and 12 ms are not remarkably different. 
A more detailed discussion regarding the different non-linear dynamics of the no-ELM and pellet-triggered ELM responses to pellet injection is included in Chapter~\ref{compare-regimes}.  
In Section~\ref{size}, we investigate the influence of the pellet size onto the observed lag-time. 

\section{Lag-time dependency on the pellet size}\label{size}

While the entire Section~\ref{timing_scan} dealt with the pellet-triggering for a medium-sized pellet of ${0.8\times 10^{20}~\mathrm{D}}$ atoms, we are now turning to a small pellet of ${0.4 \times 10^{20}~\mathrm{D}}$ atoms and a large pellet of ${1.5 \times 10^{20}~\mathrm{D}}$ atoms to investigate the dependency of the plasma response on the pellet size. The injection speed remains at $560$~m/s for all the cases presented in this section. Note here, that the small pellet contains less deuterium atoms than the smallest experimentally achievable pellet sizes (even after taking into account losses in the guide tube). Simulations with this unrealistically small pellet size are included here only to have smaller and larger pellet sizes than the medium-sized pellet (${0.8\times 10^{20}~\mathrm{D}}$ atoms) configuration we are studying. Small pellet injections are performed at 8 ms, 10 ms, and 12 ms. An ELM crash is not triggered in any of these cases such that we do not present a detailed analysis here.

\begin{figure}
\centering
\begin{tabular}{c c}
  \includegraphics[width=0.5\textwidth]{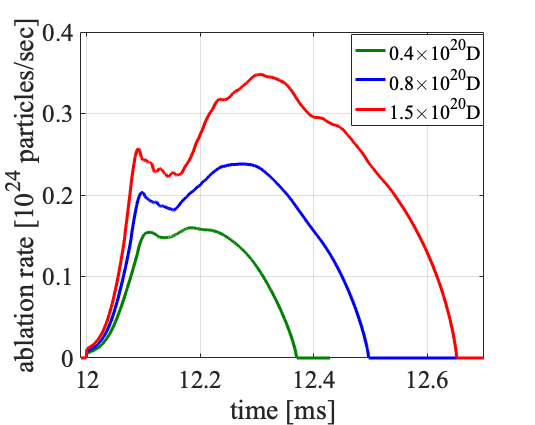} &
  \includegraphics[width=0.5\textwidth]{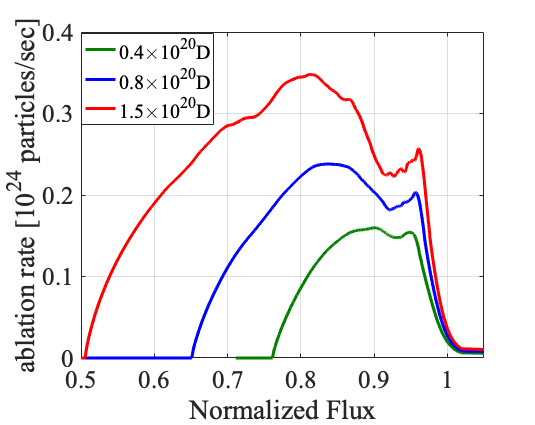} 
\end{tabular}
\caption{The pellet ablation rate versus (left panel) time and versus (right panel) normalized poloidal magnetic flux for three pellet sizes, $1.5 \times 10^{20} D$, $0.8 \times 10^{20} D$ and $0.4 \times 10^{20} D$, injected at 12 ms.}
\label{fig:ablation_12ms}
\end{figure}

Figure~\ref{fig:ablation_12ms} shows the pellet ablation rate versus time and versus normalized flux for three pellet sizes, $1.5 \times 10^{20} D$, $0.8 \times 10^{20} D$ and $0.4 \times 10^{20} D$, injected at 12 ms. 
The pellet ablation duration increases with pellet size from $350~\mathrm{\mu s}$ for the smallest pellet to $650~\mathrm{\mu s}$ for the largest pellet. All pellet sizes which have been studied in the work penetrate beyond the pedestal top ($\Psi_N \sim 0.93$), and the large pellets can reach the core plasma, $\Psi_N \sim 0.5$. 
Note, the pellet penetration depth is not possible to be estimated by a simple function of pellet sizes. This is because the penetration depth depends on the temperature and density taken along the whole pellet trajectory, and the local temperature and the local density evolve in time not only due to adiabatic ablation, but in particular due to MHD induced transport. Depending on the excited MHD activity, e.g. a pellet-triggered ELM, the local temperature abruptly changes. 
An empirical penetration-depth scaling based on a statistical model has been studied for the HFS-launched pellets in ASDEX Upgrade~\cite{Belonohy2008}. The empirical scaling includes the ELM induced temperature and density drops as it is based on experimental penetration depth values and the scaling is performed on global plasma parameters which change weakly during the ELM crash. However, it comes to its limits when strong MHD activity modifies the temperature profile strongly during the ablation process; also the distinct differences in material assimilation between HFS and LFS injections are not part of that model. Therefore, the pellet penetration depth in realistic scenarios based on the physics process could be better obtained from non-linear MHD simulations such as JOREK.

\begin{figure}
\centering
\begin{tabular}{c c}
  \includegraphics[width=0.5\textwidth]{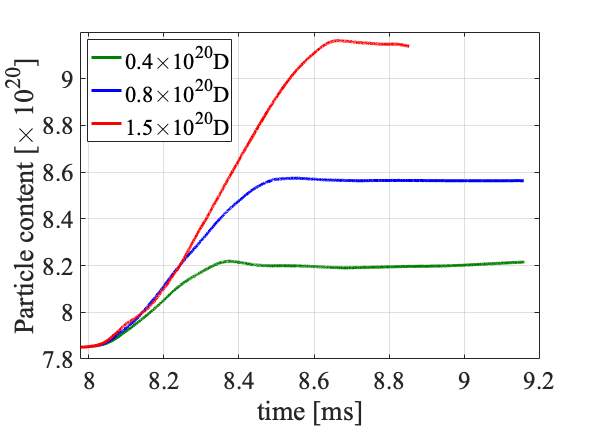} &
  \includegraphics[width=0.5\textwidth]{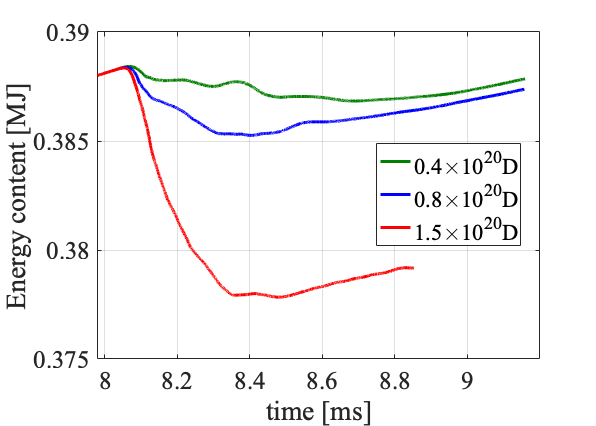} 
\end{tabular}
\caption{The time evolution of the particle and the thermal energy content inside the separatrix for three pellet sizes, ${1.5 \times 10^{20}~\mathrm{D}}$, ${0.8 \times 10^{20}~\mathrm{D}}$ and ${0.4 \times 10^{20}~\mathrm{D}}$ which are injected at 8 ms (all with 560 m/s).}
\label{fig:contents_allsize}
\end{figure}

Figure~\ref{fig:contents_allsize} shows the time evolution of the particle and the thermal energy content inside the separatrix for small, medium, and large pellets injected at 8 ms. 
The small pellet does not trigger an ELM and the particle ejection is negligible, however some of the initial pellet mass is lost in the low-temperature SOL region. The large pellet delivers 85\% of the particles into the plasma, and features a prominent drop of the energy content, meaning that an ELM has been triggered.

\begin{figure}
\centering
\includegraphics[width=0.6\textwidth]{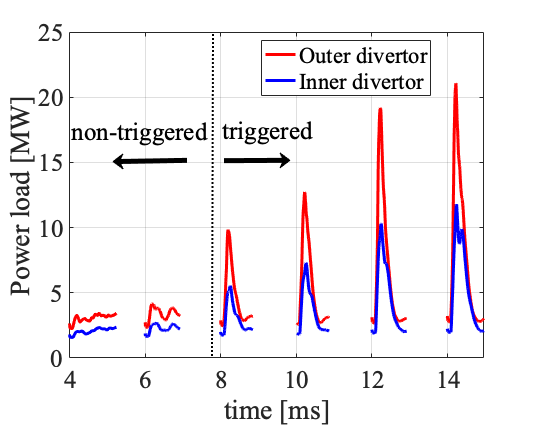}
\caption{The time evolution of the power load onto the inner and the outer divertor targets which is caused by ${1.5 \times 10^{20}~\mathrm{D}}$ pellet injections with 560 m/s of pellet injection velocity.}
\label{fig:powers_15D}
\end{figure}

Figure~\ref{fig:powers_15D} shows the power load onto the inner and the outer divertor targets which is caused by large pellet injections with 560 m/s. There is a clear transition in the heat loads between the regime where no ELM is triggered (simulations with ${t_\mathrm{inj.}\leq6~\mathrm{ms}}$) and the regime where the pellet injection causes an ELM crash (${t_\mathrm{inj.}\geq8~\mathrm{ms}}$).
The duration of the peak of the power load of the pellet-triggered ELMs is roughly ${0.4~\mathrm{ms}}$ and therefore very similar to the medium-sized pellet ELM triggering. 

The large pellet triggers an ELM that displays a toroidally asymmetric heat flux deposition profile in a similar way than the medium pellet size, as discussed in Fig.~\ref{fig:heatflux_08D_12ms_toroidalangle}.
It is important to emphasize that the structure (footprint) of the heat flux profile along the toroidal angle is largely independent of the pellet size which triggers the ELM crash. The footprint of the heat flux onto the divertor target is characterized by the magnetic field configuration, which is determined by the ELM crash itself. 

Table~\ref{table:allsizes} summarizes the relative energy loss for different pellet sizes and different injection times.
\begin{table}[htbp]
\caption{The energy loss of all injections with 560 m/s performed in this work are listed. The ELM triggering cases are highlighted with blue font. Cases marked with ``--'' were not simulated.}
\centering
\begin{tabular}{|c|c|c|c|c|c|c|c|}
\hline 
pellet size & 4 ms & 6 ms & 8 ms & 10 ms & 12 ms & 14 ms & 15 ms \\
\hline
 $0.4 \times 10^{20}~\mathrm{D}$ & -- & -- & 0.33\% & 0.38\% & 0.82\% & -- & --\\ 
\hline
 $0.8 \times 10^{20}~\mathrm{D}$ &  0.27\% & 0.5\% & 0.77\% & 1.0\% & \textcolor{blue}{3.32}\% & \textcolor{blue}{4.45}\% & \textcolor{blue}{6.33}\% \\
 \hline
 $1.5 \times 10^{20}~\mathrm{D}$  & 0.88\% & 1.32\% & \textcolor{blue}{2.63}\% & \textcolor{blue}{3.52}\% & \textcolor{blue}{4.91}\% & \textcolor{blue}{6.02}\% & --\\
 \hline
\end{tabular}
\label{table:allsizes}
\end{table}

\begin{figure}
\centering
  \includegraphics[width=0.9\textwidth]{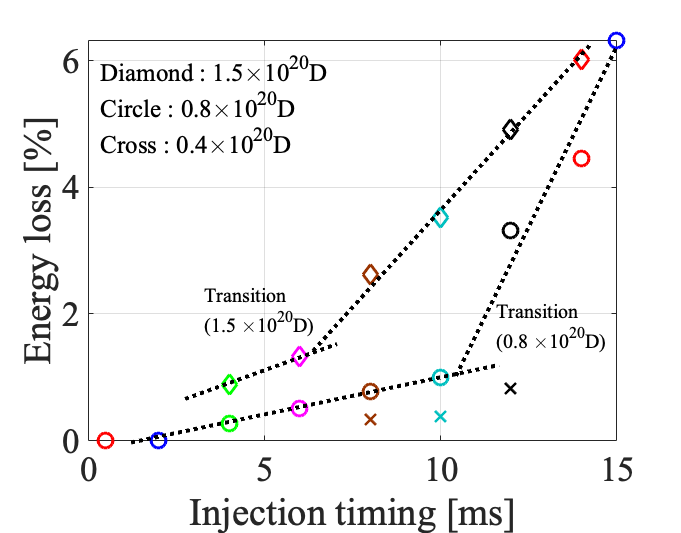} 
\caption{The relative thermal energy loss is shown for different pellet injection timings and pellet sizes. A clear transition between no-ELM and ELM triggering regimes are seen for the medium and large pellet sizes (dotted lines are shown to guide the eye).}
\label{fig:energyloss_allsize}
\end{figure}

Figure~\ref{fig:energyloss_allsize} shows the relative thermal energy loss for the different pellet injection times and different pellet sizes, i.e., the data from Table~\ref{table:allsizes}. 
The following cases show less than 1.5 \% of the energy loss, i.e. no pellet ELM triggering is achieved;
\begin{enumerate}
    \item The large pellets ($1.5 \times 10^{20}$D) injected at 4 ms and 6 ms
    \item The medium pellets ($0.8 \times 10^{20}$D) injected at 10 ms or earlier, 
    \item The small pellets ($0.4 \times 10^{20}$D) injected at all the probed injection times (very late injections were not simulated since the pellet size is anyway not experimentally relevant for ASDEX Upgrade). 
\end{enumerate}

The following cases show more than 2~\% energy loss, i.e. pellet ELM triggering is achieved;
\begin{enumerate}
    \item The large pellets ($1.5 \times 10^{20}$D) injected at, or later than, 8 ms,
    \item The base pellets ($0.8 \times 10^{20}$D) injected at, or later than, 12 ms. 
\end{enumerate}

\begin{figure}
\centering
   \includegraphics[width=0.6\textwidth]{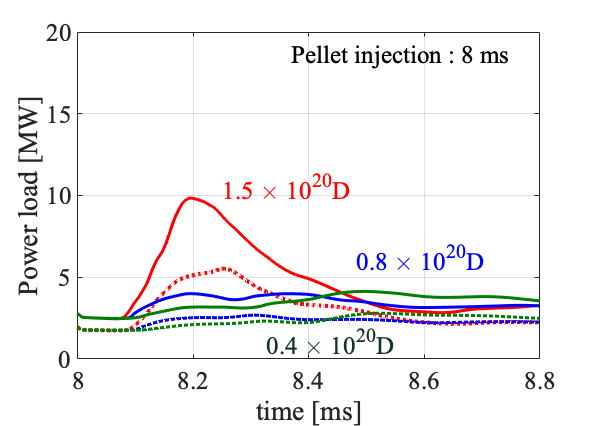}
  \\
   \includegraphics[width=0.6\textwidth]{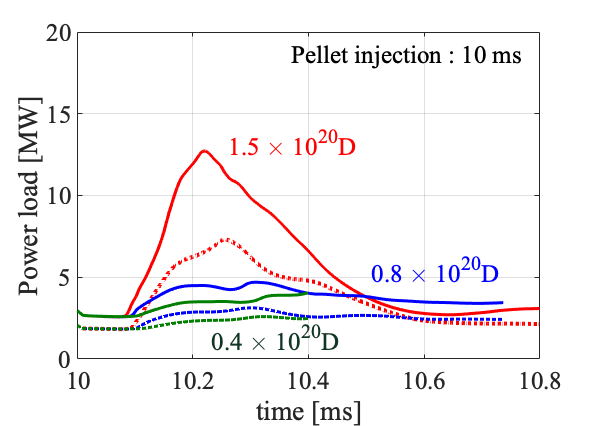}
  \\
   \includegraphics[width=0.6\textwidth]{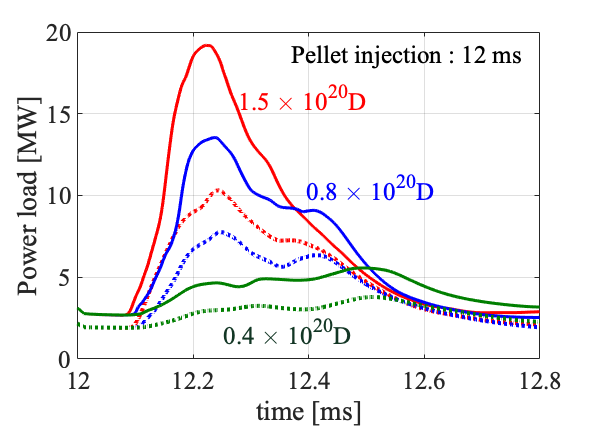}
\caption{The power load onto the divertor target. The injection of three pellet sizes $1.5 \times 10^{20} D$, $0.8 \times 10^{20} D$ and $0.4 \times 10^{20} D$ for injection times of 8 ms, 10 ms and 12 ms (keeping a fixed injection velocity of 560 m/s). The solid and the dashed lines are outer and inner divertor target, respectively.}
\label{fig:powers_allsize}
\end{figure}

Figure~\ref{fig:powers_allsize} shows the time evolution of the spatially integrated heat flux onto the divertor target versus time. All three pellet sizes ${1.5 \times 10^{20}~\mathrm{D}}$, ${0.8 \times 10^{20}~\mathrm{D}}$ and ${0.4 \times 10^{20}~\mathrm{D}}$ and injection timings of 8 ms, 10 ms and 12 ms are included. 
From the pellets injected at 8 ms and 10 ms, only the large pellet triggers an ELM and the peak of the integrated power load onto the outer divertor target reaches ${\sim 10~\mathrm{MW}}$ and ${\sim 13~\mathrm{MW}}$, respectively. Small and medium pellets do not trigger an ELM, and the integrated power load onto the outer target is ${\leq 5~\mathrm{MW}}$. For pellet injection at 12 ms, the large and medium pellets trigger ELMs and the peak of the integrated power load onto the outer divertor target reaches ${\sim 18~\mathrm{MW}}$ and ${\sim 13~\mathrm{MW}}$, respectively. At this injection time, the small pellet does not trigger an ELM. All pellet-triggered ELMs show an ELM duration of ${\sim 0.4\mathrm{ms}}$, independent from the pellet size or pellet injection time. The ELM duration is estimated here from the time of strongly increased divertor heat fluxes. Other definitions like used in the experiment (e.g., $D_\alpha$ signal) might change the time scale slightly, but are not accessible in our simulations directly.

The scan of the pellet parameters shows that the pellet size and the pellet injection time are essential parameters to determine whether pellet ELM triggering is possible. 
In this study, large pellets can trigger ELMs at an earlier time (lower pedestal pressure) with respect to the medium-sized pellets. This takes place because of the larger particle ablation leading to a stronger density/pressure perturbation which locally exceeds the MHD stability limit.
The toroidal localization of the heat flux and the radial localization of the secondary peak of the heat flux caused by pellet-triggered ELM are independent from the pellet size and pellet injection timing. 

\section{Characteristics of no-ELM and ELM-triggering response by pellet injection}\label{compare-regimes}
This section analyzes the non-linear dynamics of no-ELM and ELM-triggering responses in direct comparison based on the simulations from the previous Section. This aims to highlight why one of two regimes is entered and how they differ from each other. 

\begin{figure}
\centering
\begin{tabular}{cc}
  \includegraphics[width=0.48\textwidth]{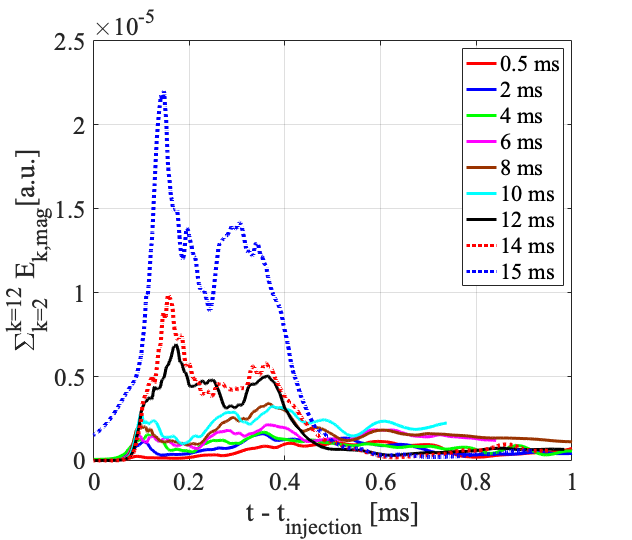} 
  &
  \includegraphics[width=0.48\textwidth]{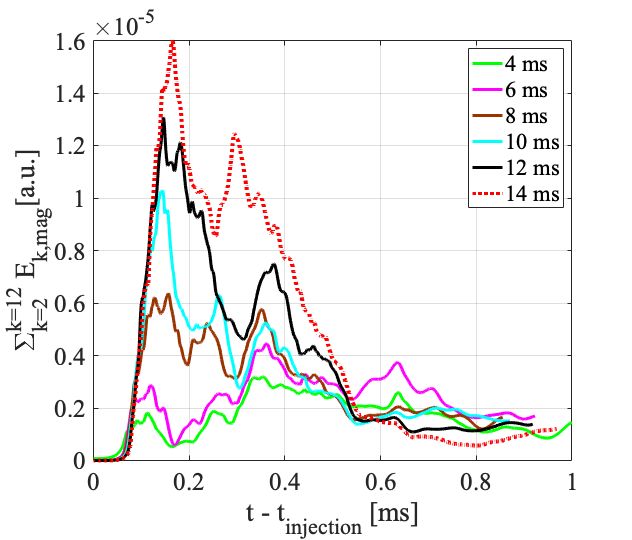} 
\end{tabular}
\caption{The time evolution of the perturbed magnetic energies in the toroidal mode numbers $n=2,3,\dots12$ is plotted for different times of pellet injection. The x-axis is shifted with respect to the injection time. The left panel contains the simulations for a pellet size of $0.8 \times 10^{20}$ atoms. The right panel corresponds to a pellet size of $1.5 \times 10^{20}$ atoms.}
\label{fig:sum_magenergies}
\end{figure}

Figure~\ref{fig:sum_magenergies} shows the time evolution of the magnetic energies corresponding to toroidal mode numbers $n=2,3,\dots12$, i.e. $\Sigma_{k=2}^{k=12} E_{k, \mathrm{mag}}$ $\equiv \Sigma_k E_\mathrm{mag}$. 
The medium-sized pellet (${0.8 \times 10^{20}~\mathrm{D}}$) injection shows a clear transition in the qualitative behaviour between the injection times 10 ms and 12 ms. This transition takes place once the injected pellet manages to trigger an ELM, i.e. ${t_\mathrm{inj.}\geq12~\mathrm{ms}}$. 
We typically find that the ELM dynamics shows a temporal sub-structure with two peaks in the magnetic energy evolution. The first steep rise of the magnetic perturbation corresponds to the onset of the ELM, and the ELM crash ends when the magnetic energy perturbation has decayed after the second peak.
The ELM onset time can be determined from the onset of the perturbed magnetic energy: ${t-t_\mathrm{inj.}\approx0.08 - 0.10~\mathrm{ms}}$.
The large pellet (${1.5 \times 10^{20}~\mathrm{D}}$) injection shows a similar transition in the integrated magnetic energy between the injection timings of 6 ms and 8 ms reflecting the ELM triggering from 8 ms onward. 
Table~\ref{table:ELMonset} summarizes the pellet-triggered ELM cases; pellet size, time of the ELM onset, pellet location at the ELM onset, and the number of pellet particles deposited at the time of the ELM onset.
Table~\ref{table:max_magenergy} summarizes the pellet-triggered ELM cases; pellet size, time of the peak of $\Sigma_k E_\mathrm{mag}$, pellet location at the peak of $\Sigma_k E_\mathrm{mag}$, and the number of pellet particles deposited at the time of the peak of $\Sigma_k E_\mathrm{mag}$.
The medium-sized ($0.8 \times 10^{20}$D) pellets injected in the later times (${t_\mathrm{inj.} \geq 12~\mathrm{ms}}$) show that the ELM triggering always occurs when the pellet is close to the pedestal top. The number of ablated atoms at the ELM onset is increasing slightly for later injection times. For the large pellet cases, the dependency on the number of ablated atoms is not observed.
\begin{table}[htbp]
\caption{The summary of pellet size, time of the ELM onset, pellet location at the ELM onset, and pellet particle deposition in the plasma up to the ELM onset}
\centering
\begin{tabular}{|c|c|c|c|c|}
\hline 
pellet size & inj. time & time of ELM onset & Pellet location & particle deposition\\
\hline
$0.8 \times 10^{20}$D & 12 ms & 12.098 ms & 0.956 & $0.078 \times 10^{20}$ \\
$0.8 \times 10^{20}$D & 14 ms & 14.099 ms & 0.952 & $0.084 \times 10^{20}$ \\
$0.8 \times 10^{20}$D & 15 ms & 15.113 ms & 0.945 & $0.089 \times 10^{20}$ \\
\hline
$1.5 \times 10^{20}$D & 8 ms & 8.084 ms & 0.965 & $0.09 \times 10^{20}$ \\
$1.5 \times 10^{20}$D & 10 ms & 10.087 ms & 0.9632 & $0.089 \times 10^{20}$ \\
$1.5 \times 10^{20}$D & 12 ms & 12.0913 ms & 0.9606 & $0.089 \times 10^{20}$ \\
$1.5 \times 10^{20}$D & 14 ms & 14.089 ms & 0.958 & $0.091 \times 10^{20}$ \\
 \hline
\end{tabular}
\label{table:ELMonset}
\end{table}
\begin{table}[htbp]
\caption{The summary of pellet size, time of the peak of $\Sigma_{k=2}^{k=12} E_{k, \mathrm{mag}}$, pellet location at the peak of $\Sigma_{k=2}^{k=12} E_{k, \mathrm{mag}}$, and pellet particle deposition in the plasma up to the time of maximum $\Sigma_{k=2}^{k=12} E_{k, \mathrm{mag}}$.}
\centering
\begin{tabular}{|c|c|c|c|c|}
\hline 
pellet size & inj. time & peak time & Pellet location & particle deposition\\
\hline
$0.8 \times 10^{20}$D & 12 ms & 12.173 ms & 0.9081 & $0.22 \times 10^{20}$ \\
$0.8 \times 10^{20}$D & 14 ms & 14.152 ms & 0.9175 & $0.1910 \times 10^{20}$ \\
$0.8 \times 10^{20}$D & 15 ms & 15.1465 ms & 0.9223 & $0.157 \times 10^{20}$ \\
\hline
$1.5 \times 10^{20}$D & 8 ms & 8.154 ms & 0.9207 & $0.2439 \times 10^{20}$ \\
$1.5 \times 10^{20}$D & 10 ms & 10.143 ms & 0.9273 & $0.217 \times 10^{20}$ \\
$1.5 \times 10^{20}$D & 12 ms & 12.146 ms & 0.9251 & $0.2189 \times 10^{20}$ \\
$1.5 \times 10^{20}$D & 14 ms & 14.1608 ms & 0.9118 & $0.2654 \times 10^{20}$ \\
 \hline
\end{tabular}
\label{table:max_magenergy}
\end{table}

\begin{figure}
\centering
  \includegraphics[width=0.95\textwidth]{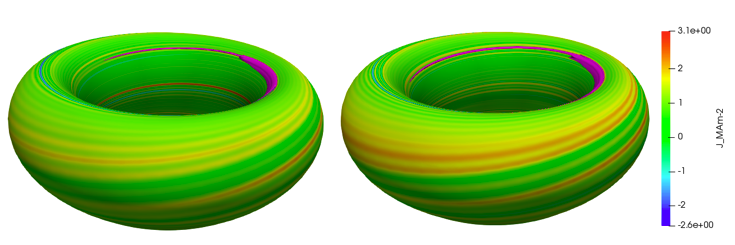} 
\caption{The $0.8 \times 10^{20}$D pellet-induced 3D pressure perturbation of 15 kPa is shown as purple band (left panel) at the time of ${\mathrm{max}(\Sigma_k E_\mathrm{mag})}$ (see Fig.~\ref{fig:sum_magenergies}, left), ${t=10.097~\mathrm{ms}}$ (${t_\mathrm{inj.}=10~\mathrm{ms}}$) and (right panel) at the ELM onset ${t=12.098~\mathrm{ms}}$ (${t_\mathrm{inj.}=12~\mathrm{ms}}$). Left panel is the pellet injection at 10 ms which does not trigger an ELM, and right panel is injection at 12 ms which triggers an ELM. The pseudocolor plot shows the current density on the flux surface of the pellet location, ${\Psi_\mathrm{N,p}=0.957}$ and ${\Psi_\mathrm{N,p}=0.956}$ for ${t_\mathrm{inj.}=10~\mathrm{ms}}$ and ${t_\mathrm{inj.}=12~\mathrm{ms}}$, respectively.}
\label{fig:3Dplot_ELMonset}
\end{figure}

The pressure perturbation of 15 kPa induced by a medium-sized pellet is shown in Fig.~\ref{fig:3Dplot_ELMonset} as a purple band together with a color contour of the current density on the pellet location at the time of ${\mathrm{max}(\Sigma_k E_\mathrm{mag})}$ for ${t=10.097~\mathrm{ms}}$ (${t_\mathrm{inj.}=10~\mathrm{ms}}$) in the left panel, and in the right panel at the ELM onset ${t=12.098~\mathrm{ms}}$ (${t_\mathrm{inj.}=12~\mathrm{ms}}$).
Current density on the magnetic flux surface of the pellet location where $\Psi_\mathrm{N,p}=0.957$ and $\Psi_\mathrm{N,p}=0.956$ for $t_\mathrm{inj.}=10~\mathrm{ms}$ and $t_\mathrm{inj.}=12~\mathrm{ms}$ is shown, respectively.
As observed in Fig.~\ref{fig:3Dplot_ELMonset}, the 3D pressure perturbation which exceeds 15 kPa shows similar profiles between 10 ms and 12 ms, however the peak values of the pressure perturbation caused by pellets are different, 31.6~kPa and 34.7~kPa for $t_\mathrm{inj.}=10~\mathrm{ms}$ and $t_\mathrm{inj.}=12~\mathrm{ms}$, respectively. The three-dimensionally localized pressure perturbation excites the current density perturbation. Therefore, larger pressure perturbation creates larger current density perturbation at the flux surface of the pellet location. The perturbations of pressure and the current density act as a seed perturbation. It is known that the seed perturbation plays an important role in the ELM crash \cite{Cathey2020}. Therefore, the small difference of the amplitude of the seed perturbation is not negligible, rather, it should be considered to play an important role.
The key parameter of pellet ELM triggering is this three-dimensional localized pressure perturbation. JOREK solves the time-evolution of the pressure, density and the current density in a self-consistent way and, therefore, the local current density is also perturbed. The localized perturbation introduces a breaking of toroidal and poloidal symmetry in MHD stability limit as observed in JOREK simulations.

\begin{figure}[h!]
\centering
  \includegraphics[width=0.9\textwidth]{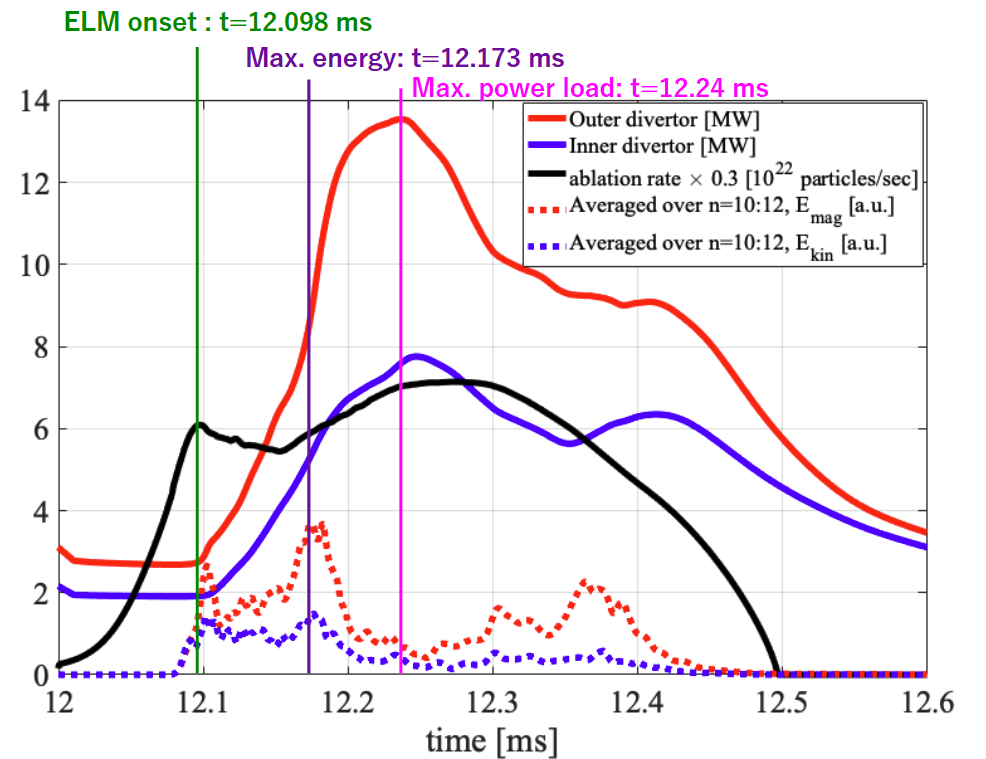} 
\caption{The time evolution of the power load onto the divertor targets, the ablation rate, and the magnetic and kinetic energies which are averaged over high toroidal modes, $n=10...12$, of the case of medium-sized ($0.8 \times 10^{20}$D) pellet injection at 12 ms. }
\label{fig:threshold_08D_12ms}
\end{figure}

It is important to emphasize that there is a delay between the ELM onset, the time of $\mathrm{max}(\Sigma_k E_\mathrm{mag})$ and the time slice of the maximum power load onto the outer target ($P_\mathrm{div,out}$). 
Figure~\ref{fig:threshold_08D_12ms} shows the time evolution of $P_\mathrm{div,in/out}$, the ablation rate, and the magnetic/kinetic energies of high toroidal modes, $n=10-12$, of the case of medium-sized ($0.8 \times 10^{20}$D) pellet injection at 12 ms.
There is a delay of ${\sim0.075~\mathrm{ms}}$ between the ELM onset (defined through the initial drop in the ablation rate) and time at ${\mathrm{max}(\Sigma_k E_\mathrm{mag})}$, and there is further delay of ${\sim0.142~\mathrm{ms}}$ between the ELM onset and the peak of $P_\mathrm{div,out}$. The delay in these events is not unexpected as it comes from the distance between separatrix and the divertor target along the magnetic field lines. The heat released from the plasma by the pellet-triggered ELM reaches the divertor target with the time scale of parallel heat diffusion along the stochastic field lines.

\begin{figure}
\centering
  \includegraphics[width=0.98\textwidth]{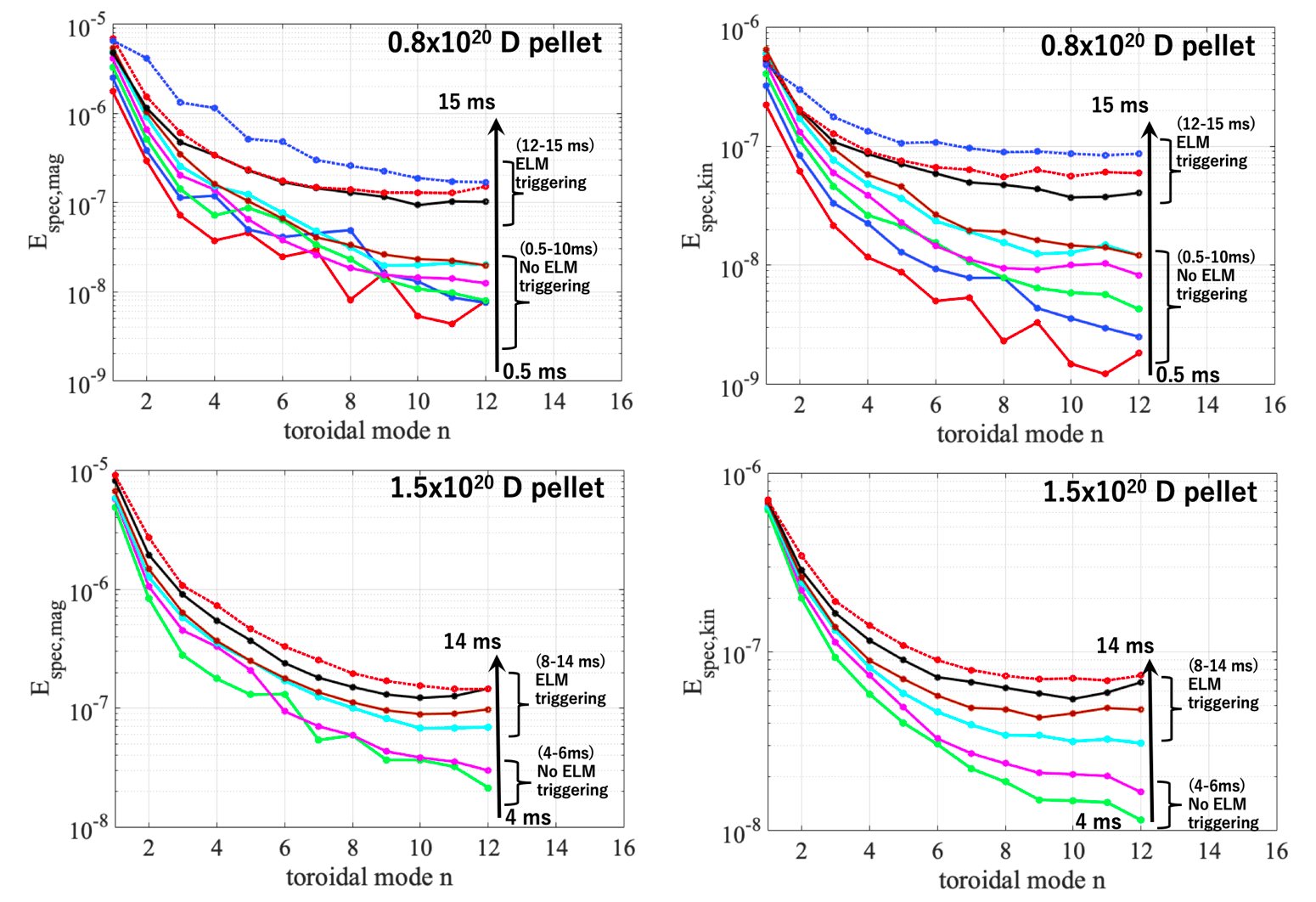}
\caption{Time-averaged toroidal spectrum over the pellet ablation process. Left panel is the magnetic spectrum and right panel is the kinetic spectrum. Top panels show $0.8\times10^{20}$D pellet, bottom panels show $1.5\times10^{20}$D pellet.}
\label{fig:spectrum_08D}
\end{figure}

Figure~\ref{fig:spectrum_08D} shows the toroidal spectrum of the kinetic and magnetic energies which are time-averaged over the pellet ablation process for the medium-sized pellet (see Table~\ref{table:08D_summary} for detailed information) and for the large pellet. 
In cases of ELM triggering of the medium-sized pellet (12 ms to 15 ms), the non-linear spectrum is significantly broader than in cases without an ELM being triggered (${t_\mathrm{inj.}\leq10~\mathrm{ms}}$). 
The large pellet cases show the broader non-linear spectrum in the pellet injections later than 8 ms.
The analysis of the toroidal spectrum is robust towards different ways of performing the time averaging. Performing the time-averaging over the duration of the ELM event (defined by the excitation of high-$n$ modes) shows the same conclusion.

\begin{figure}
\centering
  \includegraphics[width=0.98\textwidth]{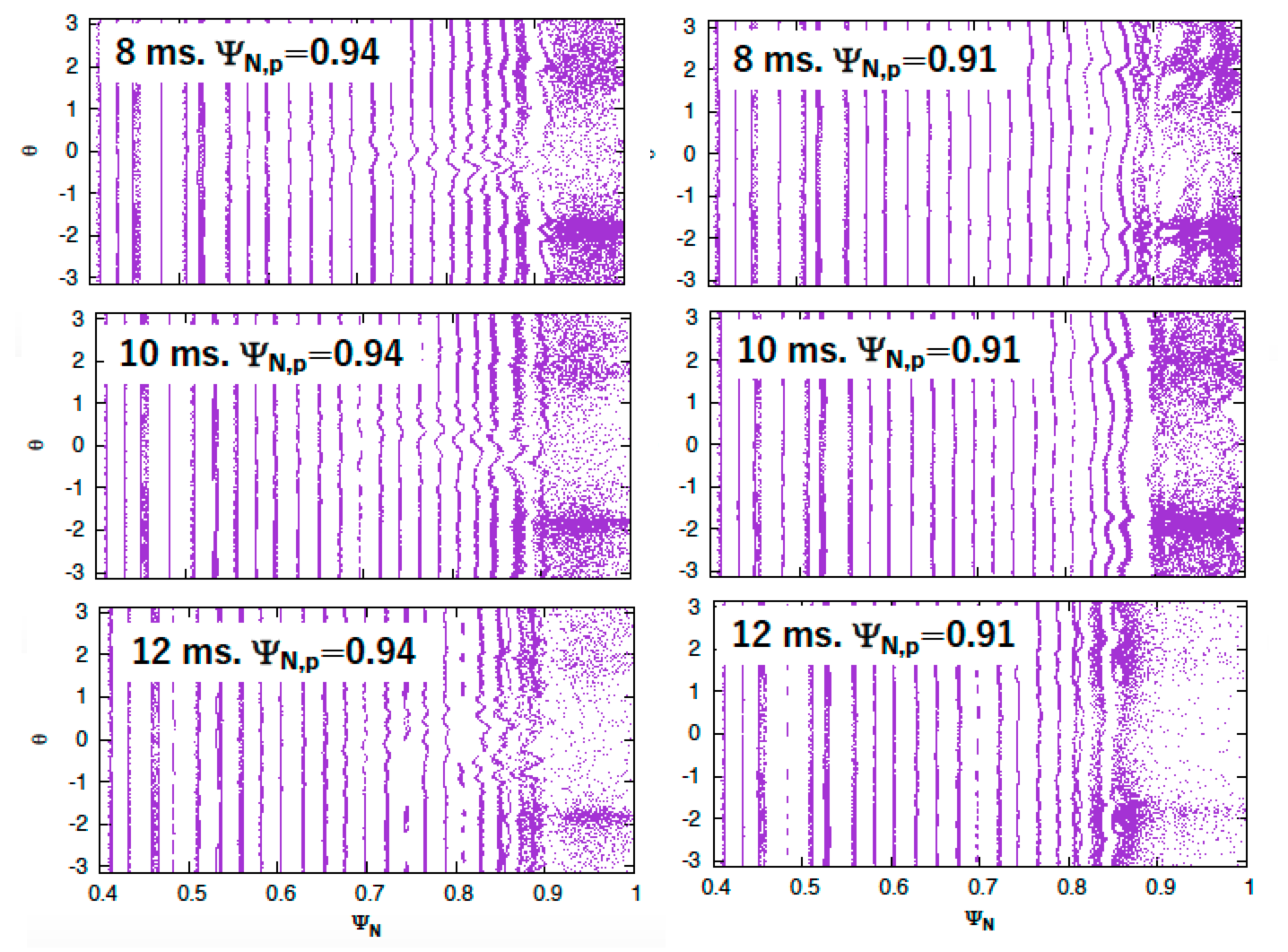} 
\caption{Poincar{\'e} plots for the pellet locations at $\Psi_{N,p}=0.94$ and $\Psi_{N,p}=0.91$, and for the injection times of $t_\mathrm{inj.} = 8~\mathrm{ms}$, 10 ms, and 12 ms of the medium-sized ($0.8\times10^{20}$D) pellet injections.} 
\label{fig:poincare_pellet_location_08D}
\end{figure}

Figure~\ref{fig:poincare_pellet_location_08D} shows the Poincar{\'e} plots for the times at which the medium-sized pellet is located at ${\Psi_\mathrm{N,p}=0.94}$ and ${\Psi_\mathrm{N,p}=0.91}$, for the injection times of ${t_\mathrm{inj.} = 8~\mathrm{ms}}$, 10 ms, and 12 ms of the medium-sized (${0.8\times10^{20}}$D) pellet injections.
Since the injection velocity and pellet trajectory is the same for all the cases considered in Fig.~\ref{fig:poincare_pellet_location_08D}, the pellet position is only dependent on ${t-t_\mathrm{inj.}}$. For each of the investigated pellet positions, ${\Psi_{N,p}=0.94}$ and 0.91, the times are ${t-t_\mathrm{inj.}=0.12~\mathrm{ms}}$ and 0.17~ms, respectively. As the pellet enters the plasma, the confining magnetic field starts to become perturbed and reconnection takes place. Therefore, a stochastic region is formed at the edge of the plasma due to the pellet-induced perturbation. On top of the pellet-induced perturbation, the response of the plasma is present. In fact, there is a visible difference in the response of the confining magnetic field between the cases where no ELM is triggered (${t_\mathrm{inj.}\leq10~\mathrm{ms}}$) and the case where the pellet-triggered ELM is present. 
While the stochastic region reaches only slightly further inwards for the ELM-triggering case, a significantly lower connection length to the divertor targets becomes visible. This can be seen from the far lower density of crossing points in the stochastic region for the late time point (pellet of $\Psi_\mathrm{N,p}=0.91$) for the $t_\mathrm{inj.}=12~\mathrm{ms}$ case. 

\section{Plasma response dependence on pellet injection velocity}\label{velocity}
While our study has so far investigated different injection times in the ELM cycle and different pellet sizes, we had kept the injection velocity fixed at 560 m/s to avoid changing several parameters at the same time. In this section, we turn now to the influence of the injection velocity. For this purpose, we take the large pellet ($1.5 \times 10^{20} D$ atoms) with 300~m/s of injection velocity injected at different times. Then we analyze the large pellet injected 8 ms and reduce the injection velocity from our reference value of 560~m/s to 300~m/s and to 240~m/s which are in the achievable range of the experiment \cite{Lang_2014}. 
We are aware that pellet size and injection velocity cannot be changed fully independently in the experiment and we will discuss this aspect in the conclusions. 

\begin{figure}
\centering
\begin{tabular}{c c}
  \includegraphics[width=0.8\textwidth]{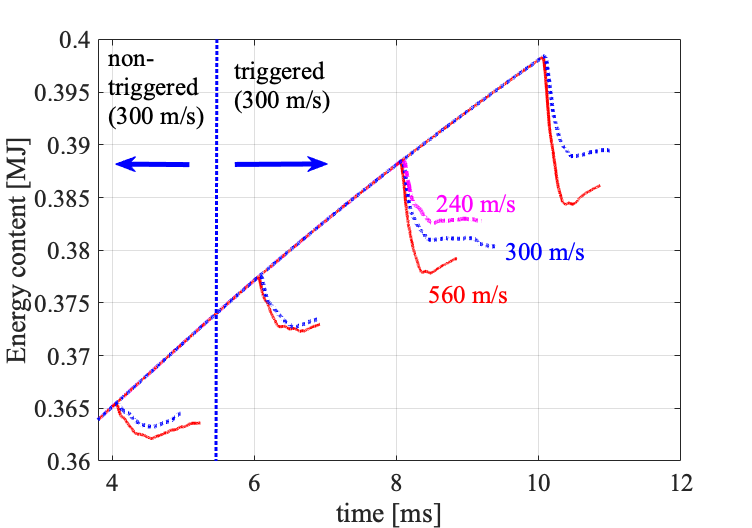} 
  \\
  \includegraphics[width=0.8\textwidth]{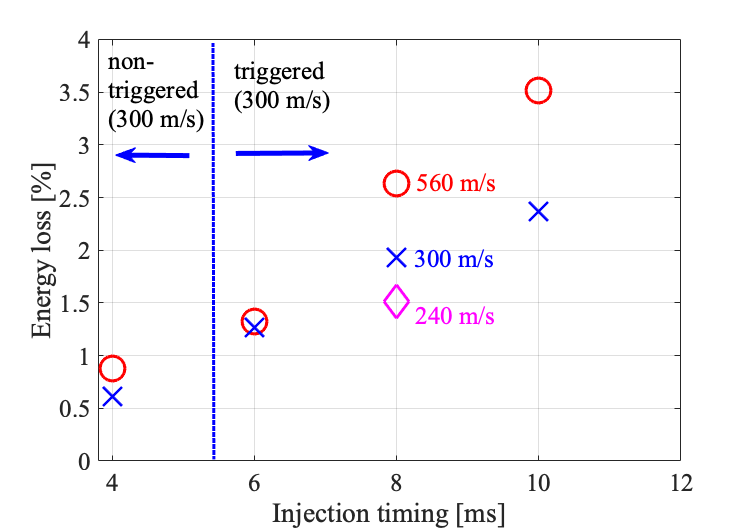} 
\end{tabular}
\caption{(Top panel) Thermal energy content inside the separatrix versus time for several different injection times and pellet velocities of 560 m/s (solid), 300 m/s (dotted) and 240 m/s (dashed). (Bottom panel) Relative loss of thermal energy for different injection times. The pellet size is $1.5 \times 10^{20} D$.}
\label{fig:energy_content_300v}
\end{figure}

Large pellet injections with different injection velocities at different time points are analyzed as shown in Fig.~\ref{fig:energy_content_300v}. With a pellet velocity of 300~m/s, injection times 4, 6, 8 and 10~ms have been simulated. For 240~m/s, only one simulation with injection at 8~ms is produced. The results for 560~m/s from the previous section are also included for reference. In particular, the thermal energy time trace and the relative energy lost as a result of the pellet injection are shown in the top and bottom figures, respectively. 
For 300~m/s injection velocity, the relative energy loss for the injection time ${t_\mathrm{inj.} = 4~\mathrm{ms}}$ is 0.61~\%, which is smaller than the relative energy loss of 0.88~\% caused by the pellet of 560~m/s velocity at the same injection time. 
The relative energy loss at ${t_\mathrm{inj.} = 6~\mathrm{ms}}$ is similar between 560 m/s and 300 m/s pellet injection velocity, 1.32~\% and 1.27~\%, respectively.
The slow pellet injections (300~m/s and 240~m/s) at 8~ms trigger an ELM albeit with lower relative energy losses (1.93~\% and 1.51~\%, respectively) than the 2.63~\% lost by the pellet injected at 560~m/s. 
The measurement of the thermal energy loss for 300~m/s injection velocity cases does not feature such a clear transition between no-ELM and ELM-triggering as clear as the one observed for ${v_\mathrm{p}=560~\mathrm{m/s}}$.
However, observing the time traces of the thermal energy content for the 560~m/s injection velocity, it can be seen that the loss rate of thermal energy is very different between the no-ELM and triggering cases as seen from the different temporal gradients of the curves. 
The analyses of the toroidal mode spectrum and the power load incident onto the divertor targets are discussed in the next paragraphs. Through these it was possible to identify that the transition from no-ELM to ELM triggering for the ${v_\mathrm{p}=300~\mathrm{m/s}}$ scenario takes place between the injection times of 4 and 6~ms (as indicated in Fig.~\ref{fig:energy_content_300v}).

\begin{figure}
\centering
  \includegraphics[width=0.98\textwidth]{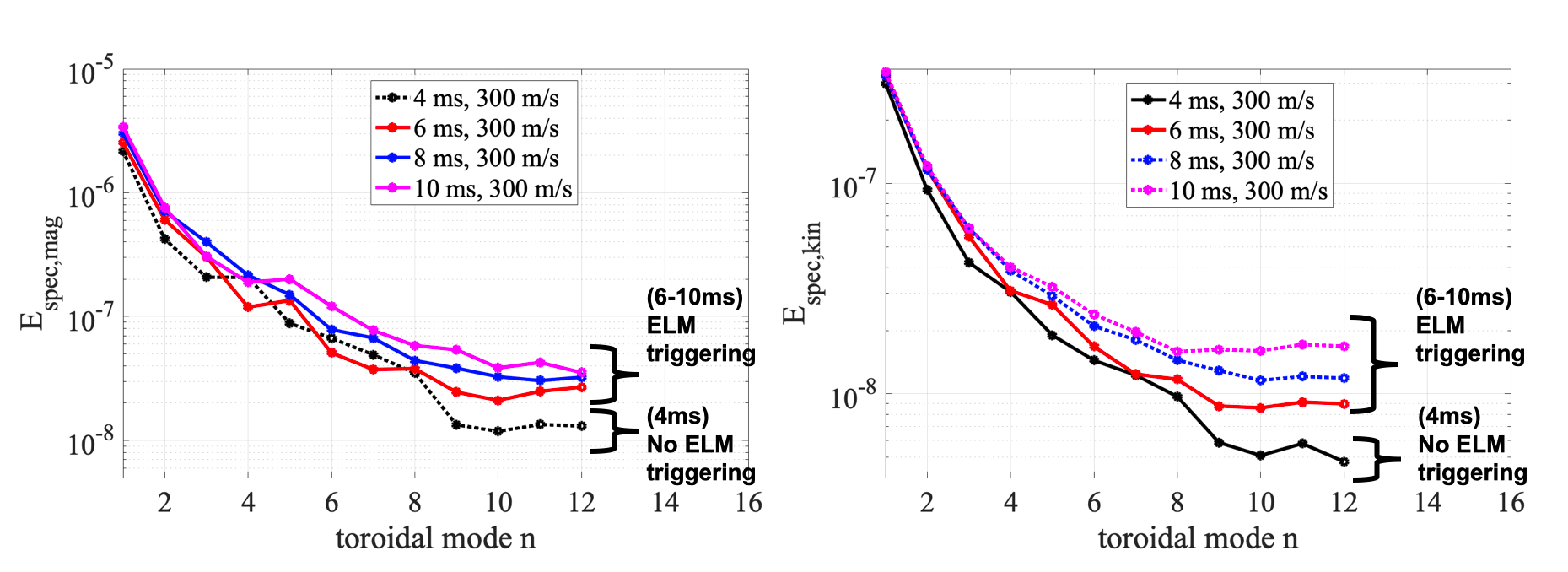}
\caption{Time-averaged toroidal spectrum over the pellet ablation process. Left panel is the magnetic spectrum and right panel is the kinetic spectrum. The large pellets with injection velocity 300 m/s are injected at 4 ms, 6 ms, 8 ms and 10 ms.}
\label{fig:spectrum_300v}
\end{figure}

Figure~\ref{fig:spectrum_300v} shows the toroidal spectrum of the magnetic and kinetic energies which are time-averaged over the pellet ablation process for the large pellet injected with the velocity of 300 m/s. 
A broadening of the non-linear spectrum is observed between pellet injection at 4~ms and at 6~ms for the pellet injection velocity of 300 m/s. 
Comparatively, the broadening of the non-linear spectrum for pellet injection at 560 m/s takes place between 6 ms and 8 ms as shown in Fig.~\ref{fig:spectrum_08D}. 
The analysis of the toroidal spectrum confirms that the timing of ELM triggering with the injection velocity is between 4 ms and 6 ms (earlier than the case of injection velocity of 560 m/s which is between 6 ms and 8 ms).

\begin{figure}
\centering
\begin{tabular}{c c}
  \includegraphics[width=0.8\textwidth]{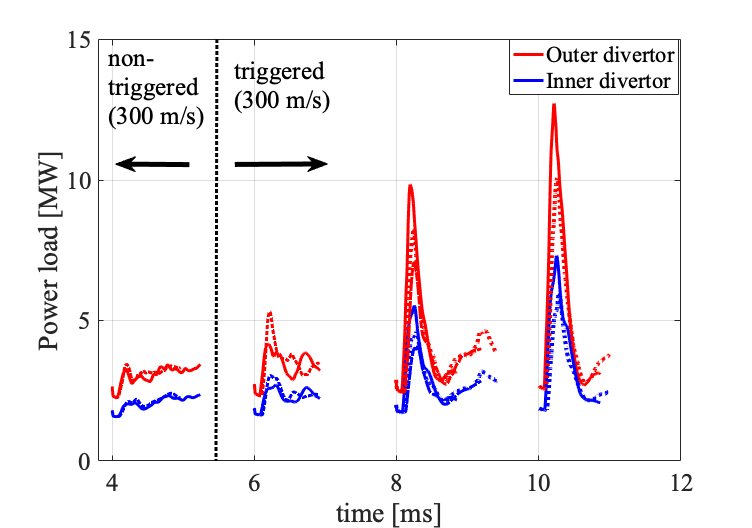} 
\end{tabular}
\caption{Time evolution of the power load onto the inner and outer divertor targets caused by $1.5 \times 10^{20} D$ pellet injections with the velocities  560 m/s (solid), 300 m/s (dotted) and 240 m/s (dashed).}
\label{fig:powers_300v}
\end{figure}

Figure~\ref{fig:powers_300v} shows the power load onto the inner and outer divertor targets caused by pellets injected with the different velocities. In this analysis, the no-ELM (4 ms) and ELM triggering cases (6 ms and later) can be distinguished for 300~m/s injection velocity since the ELM crash leads to a pronounced spike in the target power load. All considered injection velocities exhibit the lag-time during which ELM triggering is not possible, and the ELM energy losses are reduced at lower injection velocities. 
It is interesting to note, however, that the transition from no-ELM to ELM-triggering for ${v_\mathrm{p}=300~\mathrm{m/s}}$ is less clear than the transition for the faster injection velocity.

\begin{figure}
\centering
\begin{tabular}{c c}
  \includegraphics[width=0.5\textwidth]{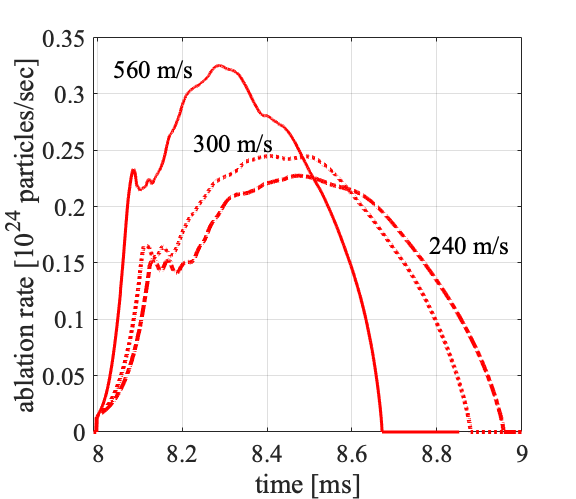} &
  \includegraphics[width=0.5\textwidth]{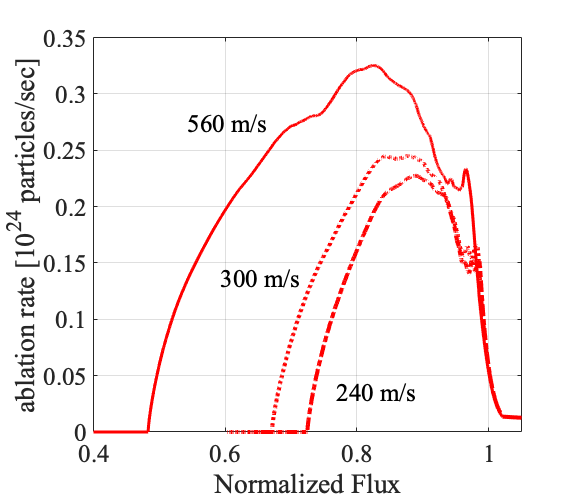} 
\end{tabular}
  \includegraphics[width=0.5\textwidth]{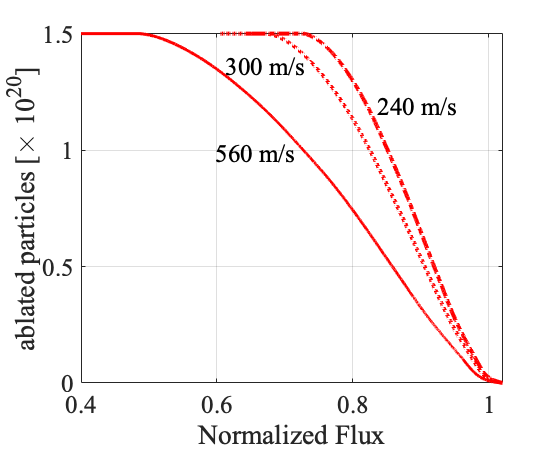} 
\caption{(Top panels) The pellet ablation rate versus time and the ablation rate profile versus normalized flux, and (bottom panel) the ablated particles versus normalized flux. The pellet size is $1.5 \times 10^{20} D$ injected at 8 ms for three injection velocity; 560 m/s (reference case), 300 m/s and 240 m/s. }
\label{fig:ablation_15D_8ms_velocity}
\end{figure}

Knowing that the injection at $t_\mathrm{inj.}=8$~ms triggers an ELM for all considered pellet velocities, this injection time is analyzed further in detail in the following.
Figure~\ref{fig:ablation_15D_8ms_velocity} shows the pellet ablation rate versus time, and the ablation rate profile versus normalized flux for three pellet injection velocities, 560~m/s, 300~m/s and 240~m/s. The reference pellet ($v_\mathrm{p}=560$ m/s) quickly reaches the high-temperature region and, therefore, the amplitude of the ablation rate is initially larger than for the cases with slower pellet injection. As the pellet which is injected with 560~m/s ablates more material in the first ${\sim0.4~\mathrm{ms}}$ than the slower pellet injections (as shown in Fig.~\ref{fig:ablation_15D_8ms_velocity}), the duration of the ablation time is shorter, but the pellet penetrates deeper than the slower injections. The cases of slower pellet injection stay in the pedestal, i.e. lower temperature region for a longer time. Therefore the pellet ablation rate is lower than the fast injection cases and penetration is not as deep. 

\begin{figure}
\centering
  \includegraphics[width=0.48\textwidth]{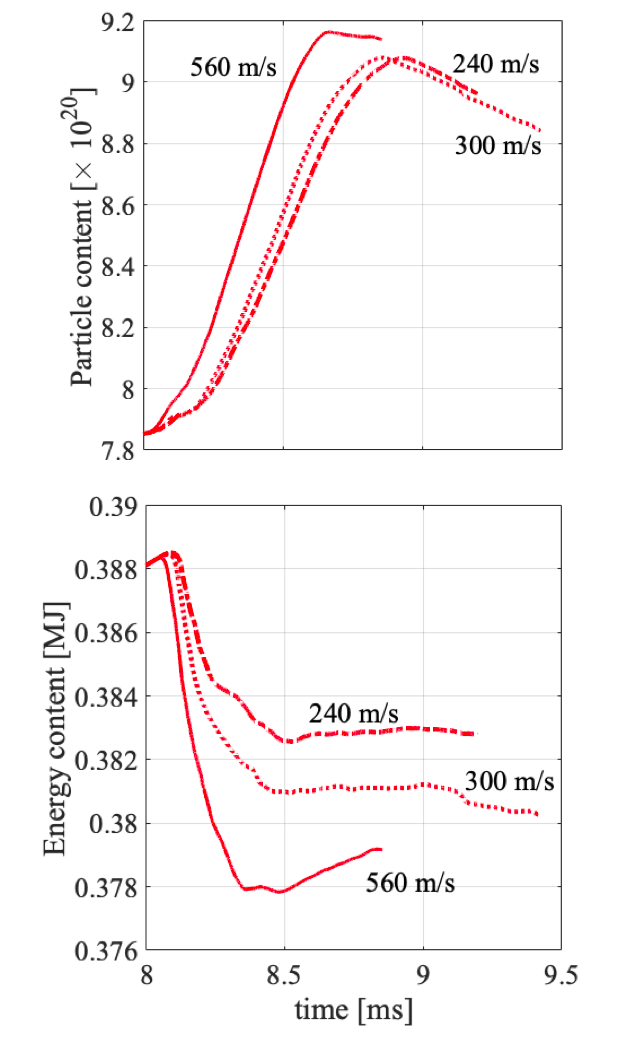} 
\caption{The particle and the energy content is plotted versus time. The pellet size is $1.5 \times 10^{20} D$ injected at 8 ms for three injection velocities; 560 m/s (reference case), 300 m/s and 240 m/s.}
\label{fig:contents_15D_8ms_velocity}
\end{figure}

Figure~\ref{fig:contents_15D_8ms_velocity} shows the time evolution of the particle and the energy content. 
After reaching the peak of the particle content, the slow injection speeds, 300~m/s and 240~m/s show a drop of the particle content in the plasma. 
The plot of the energy content shows that the fast pellet injection (560 m/s) induces a large energy drop in a short time, ${\sim 0.3~\mathrm{ms}}$. On the other hand, the slower pellet injections, 300~m/s and 240~m/s, show comparatively smaller drops of the energy content, i.e., the pellet-triggered ELM energy losses are increasing (in this case) with the injection velocity.
The slower pellet injections reach the maximum pellet ablation rate at $\Psi_N = 0.9$ which is close to the pedestal region compared to the reference case (at $\Psi_N = 0.83$), as shown in Fig.~\ref{fig:ablation_15D_8ms_velocity}. 
The region over which the pellet particles are deposited and the duration of the pellet ablation cause observable differences in terms of the duration of the energy and the particle losses caused by the pellet-triggered ELM.
For the range of injection velocities studied with large pellets, it is observed that the fast pellet injection reaches the high temperature region quickly. Therefore the ablation rate (which increases strongly with increasing temperature), becomes large. The large ablation rate related to the fast injection deposits particles deeper inside the plasma causing larger density/pressure perturbations at these locations. As a consequence, the drop of the energy content caused by fast pellet injections becomes larger and sharper compared to the slow injections.

\begin{figure}
\centering
  \includegraphics[width=0.8\textwidth]{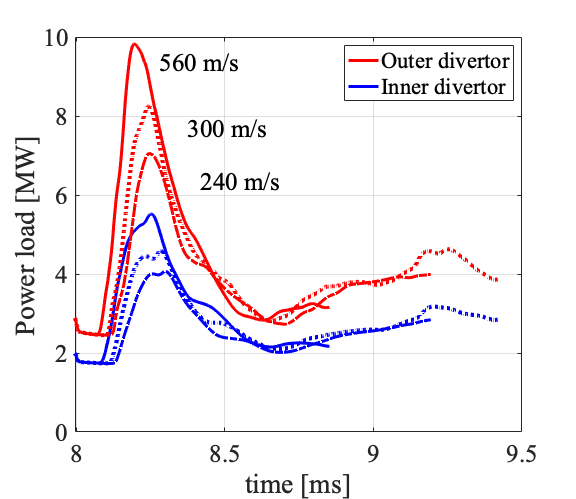} 
\caption{The time evolution of the power load onto the inner and the outer divertor targets which is caused by $1.5 \times 10^{20} D$ pellets injected at 8 ms, for 560 m/s (reference case), 300 m/s and 240 m/s.}
\label{fig:powers_15D_8ms_velocity}
\end{figure}

Figure~\ref{fig:powers_15D_8ms_velocity} shows the power load onto the inner and the outer divertor targets which is caused by $1.5 \times 10^{20} D$ pellets injected at 8 ms, for 560 m/s (reference case), 300 m/s and 240 m/s. 
The peak of the power load onto the divertor targets is increasing with the pellet injection velocity. As discussed in the previous paragraph, the thermal energy loss increases with increasing the pellet injection velocity in the present scans. The power load onto the divertor target is proportional to the thermal energy loss from the plasma.
The ELM duration is about 0.4 ms independently of the pellet injection velocity. 

The large pellet injection with 560~m/s at 8~ms triggers an ELM when the pellet location is ${\Psi_N = 0.964}$. 
The slower injections, 300 m/s and 240 m/s, at 8~ms show the ELM onset at ${\Psi_N = 0.981}$ and ${\Psi_N = 0.983}$, respectively. As shown in Fig.~\ref{fig:powers_15D}, the no-ELM cases ($t_\mathrm{inj.} = 4~\mathrm{ms}$ and 6~ms with ${v_\mathrm{p}=560~\mathrm{m/s}}$) show the power incident on the outer divertor target is less than 4.2~MW, while the ELM triggering by later pellet injection shows $ \sim 9.8~\mathrm{MW}$ (${t_\mathrm{inj.}=8~\mathrm{ms}}$ with ${v_\mathrm{p}=560~\mathrm{m/s}}$). 
The slower pellet injections trigger an ELM, but the ELM sizes are smaller than for the reference pellet injection velocity.
The experimental study of Ref~\cite{Kocsis2007} observed that the location of the seed perturbation was the same for all pellet velocities (middle of the pedestal) at the time of the ELM triggering. However, it was shown that due to the intrinsic delay of the detection, the pellet position at the ELM onset was velocity dependent, i.e., faster pellets were located deeper inside the plasma. It is worth noting, that in the way we determine the ELM onset, i.e., by the drop of the ablation rate due to the local drop of the temperature, no delay is expected such that the ELM onset time is known exactly from the simulations. Thus, the observation seems well in line with the simulation results obtained in this paper.
Note, that we vary pellet size and injection velocity independently in the simulations, while both are typically correlated in the experiment.

\begin{figure}
\centering
\begin{tabular}{c c}
  \includegraphics[width=0.5\textwidth]{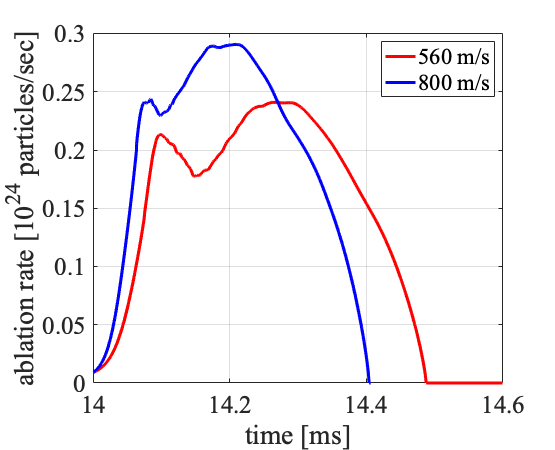}  &
  \includegraphics[width=0.5\textwidth]{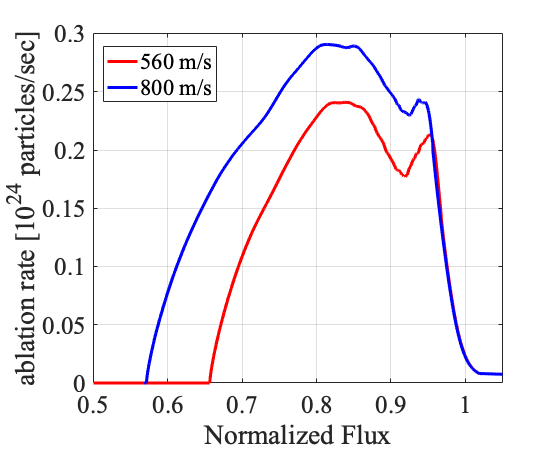} 
\end{tabular}
  \includegraphics[width=0.5\textwidth]{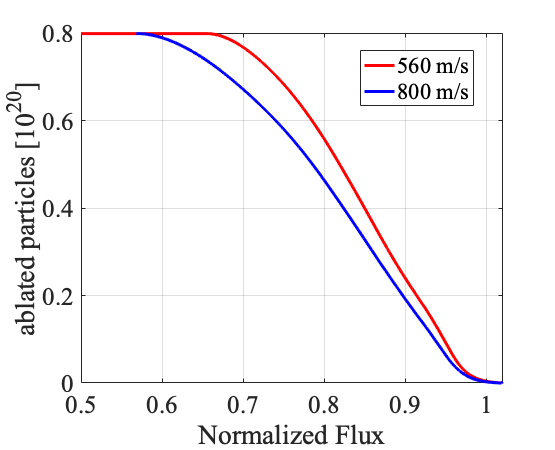} 
\caption{(Top panels) The pellet ablation rate versus time and the ablation rate profile versus normalized flux, and (bottom panel) the ablated particles versus normalized flux. The pellet size is $0.8 \times 10^{20}$D and the injection timing is 14 ms which is triggers an ELM. Red lines are 560 m/s and blue lines are 800 m/s. }
\label{fig:ablation_08D_14ms_velocity}
\end{figure}

The pellet injection velocity dependence has additionally been studied with the plasma which is very close to produce the natural ELM. The pellet of ${0.8 \times 10^{20}~\mathrm{D}}$ size is injected at 14~ms with two injection velocities: 560~m/s (reference case) and 800~m/s.
Figure~\ref{fig:ablation_08D_14ms_velocity} shows the time evolution of the pellet ablation rate and the pellet ablation rate versus normalized flux for the reference pellet and for the pellet with  ${v_\mathrm{p}=800~\mathrm{m/s}}$. The fast pellet reaches the high-temperature region quickly, therefore the pellet ablation rate increases relative to the reference case. As the pellet ablation rate is high, the fast pellet reaches the full ablation quicker than 560 m/s injection case. The fast pellet injection penetrates deeper into the plasma, ${\Psi_N \sim 0.55}$, while the reference case reaches ${\Psi_N \sim 0.65}$. 

\begin{figure}
\centering
  \includegraphics[width=0.6\textwidth]{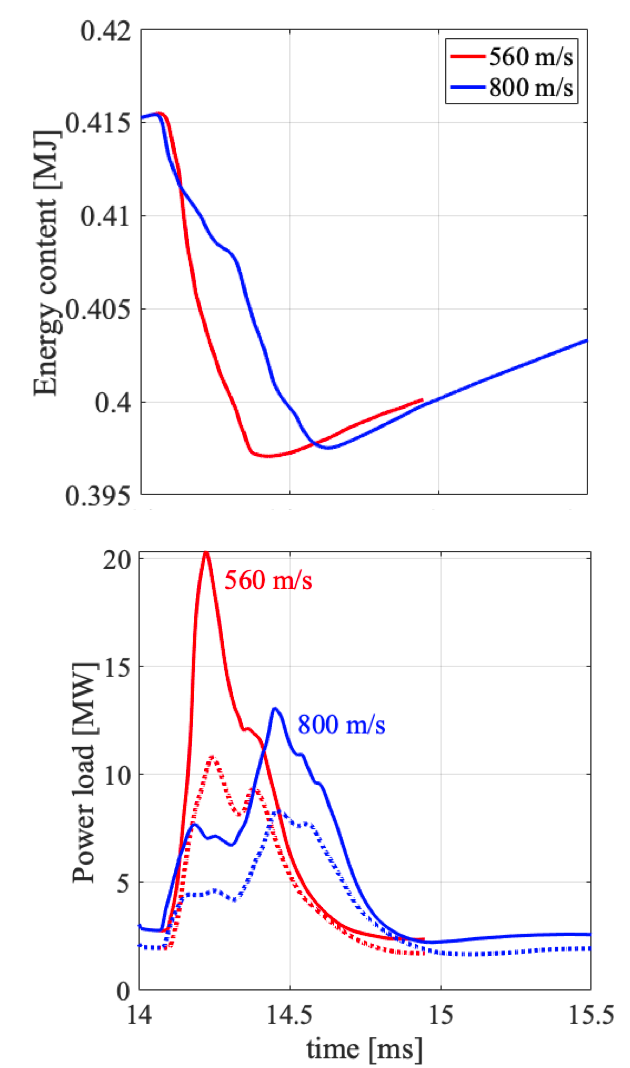}  
\caption{(Top panel) The time evolution of the energy content inside the separatrix for 560 m/s (reference injection speed) and 800 m/s. (Bottom panel) The time evolution of the power load onto the divertor targets. Solid and dashed lines are outer and inner divertor targets.}
\label{fig:powers_08D_14ms_velocity}
\end{figure}

Figure~\ref{fig:powers_08D_14ms_velocity} shows the time evolution of the energy content inside the separatrix and the power load onto the divertor targets. 
The reference case induces a sharper drop of the energy content compared to the faster pellet injection. As consequence, the power load onto the divertor target for the pellet injection with 560~m/s leads to a larger peak power load with respect to the faster pellet injection. 
The amount of the energy lost after the pellet injections with 560~m/s and 800~m/s is similar, 20.4~kJ and 18.2~kJ, respectively. This observation is in some contrast to the $v_\mathrm{p}$ scan with pellets of ${1.5 \times 10^{20}~\mathrm{D}}$ atoms injected at 8 ms, where injection at the reference velocity lead to larger losses than at slower $v_\mathrm{p}$.

\begin{figure}
\centering
  \includegraphics[width=0.8\textwidth]{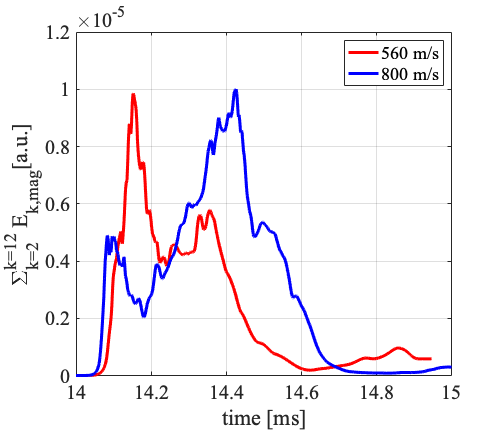} 
\caption{The time evolution of the integrated magnetic energies over $k=2-12$, $\Sigma_{k=2}^{k=12} E_{k, \mathrm{mag}}$, for 560 m/s pellet injection (reference case) and 800 m/s.}
\label{fig:mag_energies_08D_14ms_velocity}
\end{figure}

For the ${0.8 \times 10^{20}~\mathrm{D}}$ atoms pellet, injection with ${v_\mathrm{p}=560}$ and with 800~m/s cause similar ELM induced thermal energy losses. However, while the reference case causes a very fast crash within approximately ${250~\mathrm{\mu s}}$, the drop of the thermal energy is slower for the case with ${v_\mathrm{p}=800~\mathrm{m/s}}$ and it is divided into two separate energy drops. This appears to be linked to the lower material ablation in the pedestal region for the fast injection case (see Fig.~\ref{fig:ablation_08D_14ms_velocity}). Figure~\ref{fig:mag_energies_08D_14ms_velocity} shows the time evolution of ${\Sigma_{k=2}^{k=12} E_{k, \mathrm{mag}}\equiv\Sigma_k E_\mathrm{mag}}$ for the reference case and for the ${v_\mathrm{p}=800~\mathrm{m/s}}$ case. 
The reference case shows the peak of ${\Sigma_k E_\mathrm{mag}}$ at 14.15~ms, which is much earlier than the time of full ablation, 14.488~ms. The pellet excites the ELM during the pellet ablation process. 
On the other hand, the pellet injection of 800~m/s shows the peak of ${\Sigma_k E_\mathrm{mag}}$ at 14.42~ms which is after the time of full ablation, 14.40~ms.

\begin{figure}
\centering
  \includegraphics[width=0.98\textwidth]{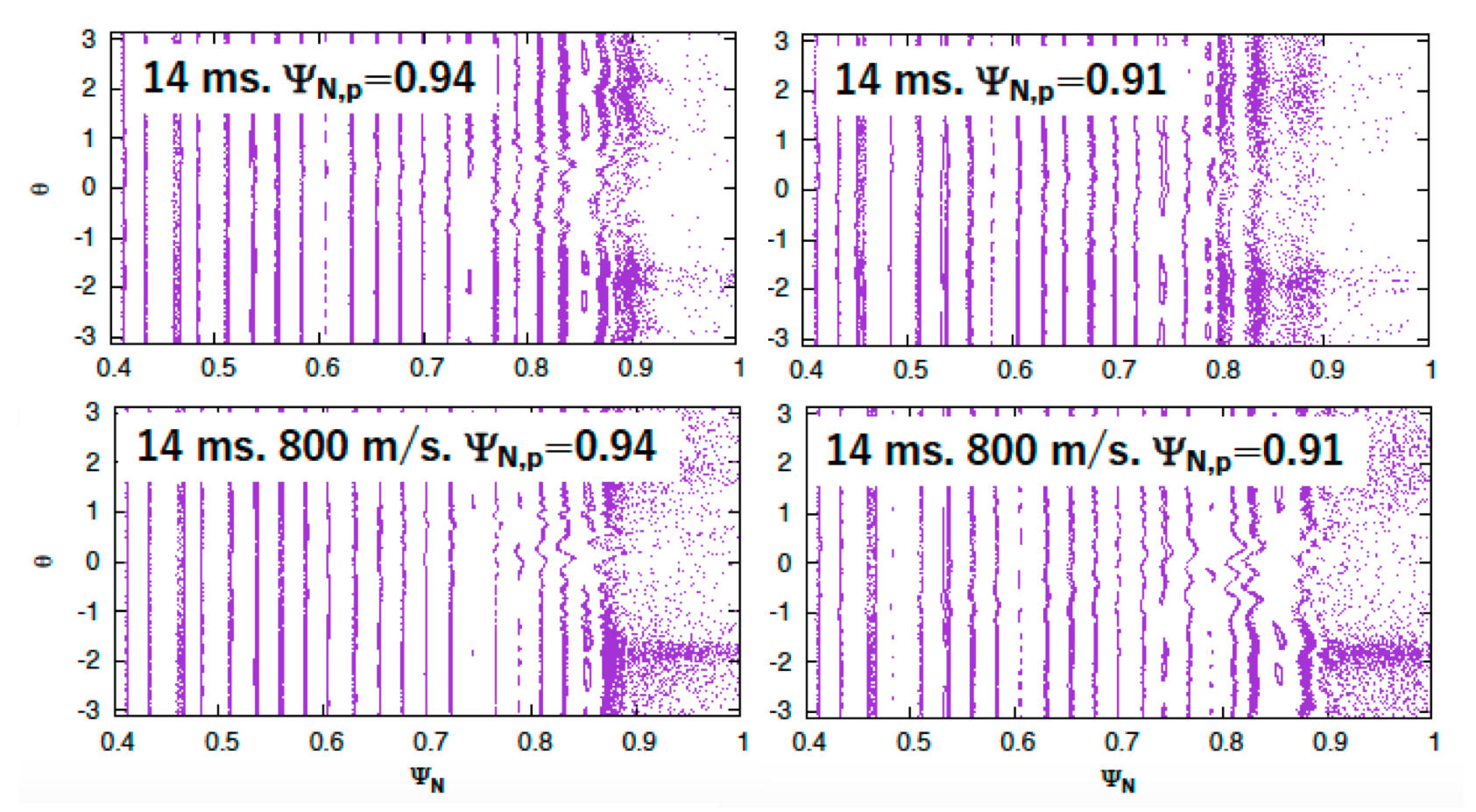} 
\caption{Poincar{\'e} plots for the pellet locations at $\Psi_{N,p}=0.94$ and $\Psi_{N,p}=0.91$, for the injection velocities of 560 m/s and 800 m/s. }
\label{fig:poincare_velocity_14ms}
\end{figure}

Figure~\ref{fig:poincare_velocity_14ms} shows the Poincar{\'e} plots for the pellet locations at $\Psi_{N,p}=0.94$ and $\Psi_{N,p}=0.91$, for the injection velocities of 560 m/s and 800 m/s. 
The pellet injection velocities give slightly different structures of stochastic layer although the width of the layer is comparable when the pellet location is the same. 
Comparing the Poincar\'e plots with two different injection velocities shows that the case with $v_\mathrm{p}=560~\mathrm{m/s}$ has a stochastic region with a far lower connection length. As a result, field lines from this region hit the divertor targets after a lower number of toroidal turns reflected in a lower density of points in the plot. This partly explains the stronger losses observed by the pellet injection at reference velocity.

\begin{figure}
\centering
  \includegraphics[width=0.98\textwidth]{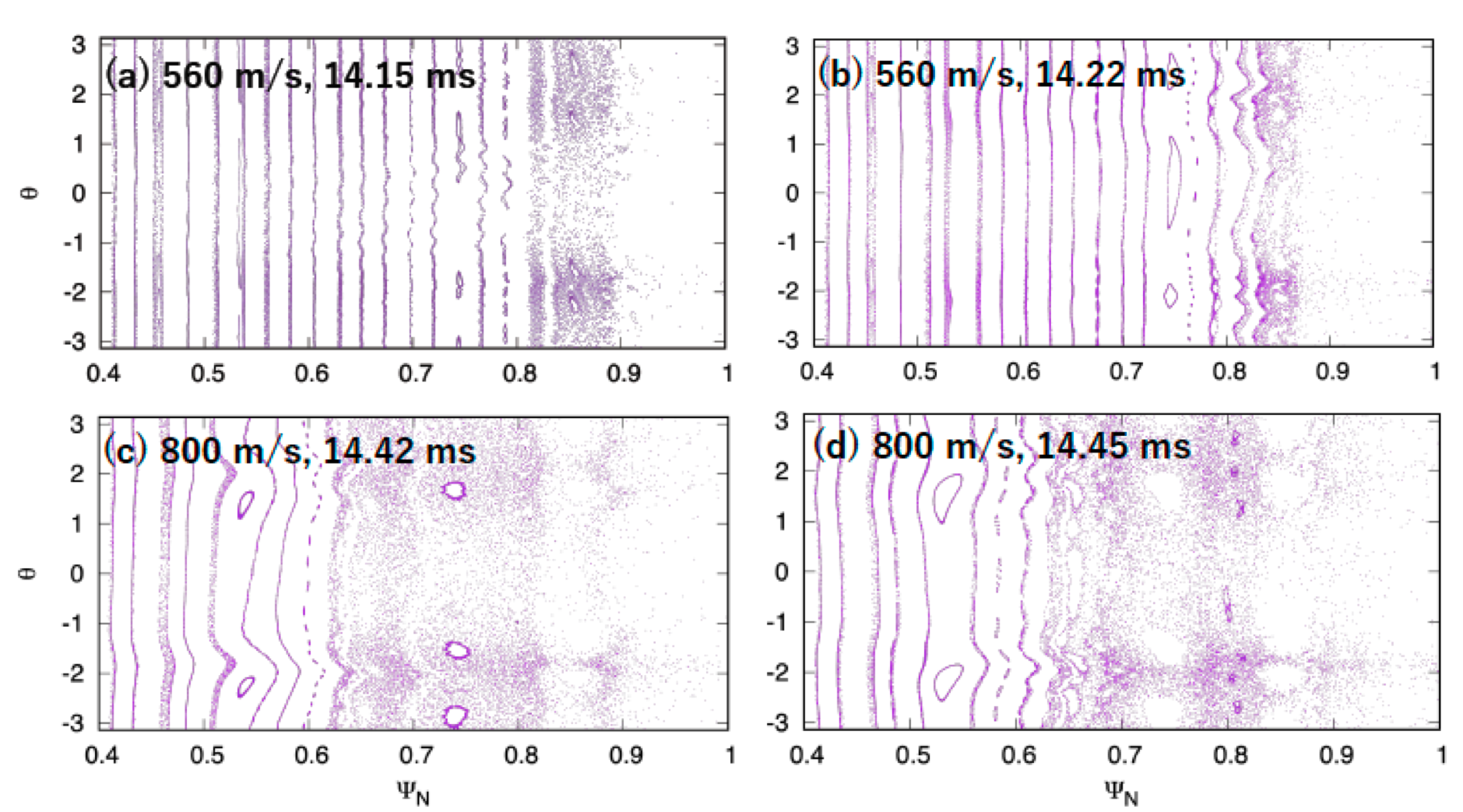} 
\caption{$0.8 \times 10^{20}$D pellet injection at 14 ms. (Left column) Poincare plot at the maximum magnetic energy $\Sigma_{k=2}^{k=12} E_{k, \mathrm{mag}}$ for pellet injection speed of 560 m/s (14.15 ms) and 800 m/s (14.42 ms). (Right column) Poincare plot at the maximum power load onto the divertor target for pellet injection speed of 560 m/s (14.22 ms) and 800 m/s (14.45 ms).}
\label{fig:poincare_08D_velocity}
\end{figure}

\begin{figure}
\centering
  \includegraphics[width=0.8\textwidth]{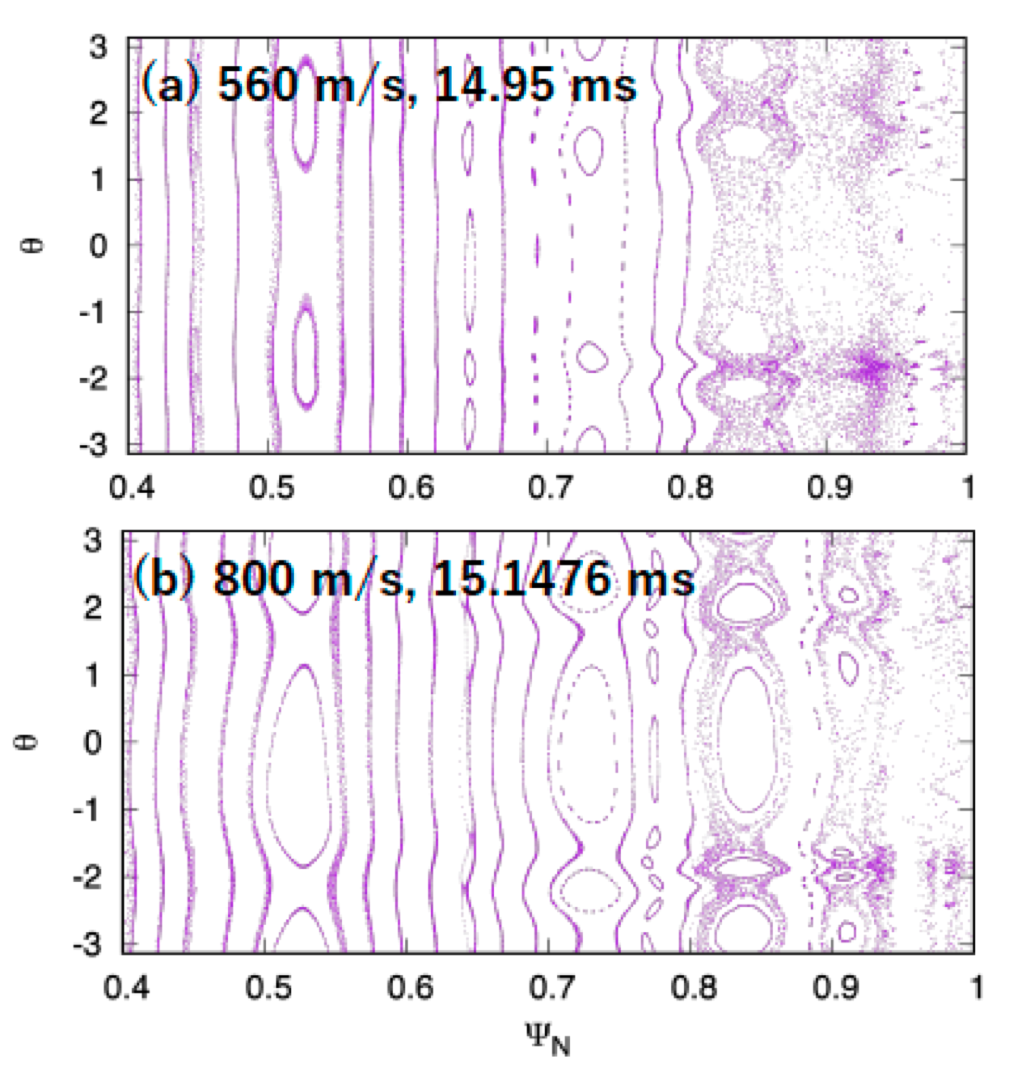} 
\caption{Poincar{\'e} plots during the relaxation state after the ELM crashes for the injection velocities of 560 m/s and 800 m/s. The times 14.95 ms (560 m/s) and 15.1476 ms (800 m/s) are plotted, which are 0.5 ms after the end of the thermal energy losses. }
\label{fig:poincare_08D_velocity_05ms}
\end{figure}

Besides the different dynamics of the ELM crash, the fast injection also excites core modes as seen in the Poincar{\'e} plots of Figure~\ref{fig:poincare_08D_velocity}.
Figure~\ref{fig:poincare_08D_velocity_05ms} shows the Poincar{\'e} plot in the relaxation state after the ELM crash for the two injection velocities. The times of 14.95 ms ($v_\mathrm{p}=$560 m/s) and 15.1476 ms (800 m/s) which are taken $0.5$ ms after the minimum of the thermal energy content. In the fast pellet injection case (800 m/s), the width of the 2/1 magnetic island is about 3 cm and the width of the 3/1 magnetic island is about 3.5 cm indicating that both might become NTMs (neoclassical tearing modes) in this scenario. In the 560 m/s injection case, the island widths are in the range of 1 cm only and therefore possibly too small for becoming NTMs. The further evolution of these core modes is beyond the scope of this work. 

\section{Conclusions and Outlook}\label{conclusions}

Non-linear MHD simulations of ELM triggering by pellet injection was studied based on an ASDEX Upgrade H-mode plasma with JOREK including realistic ExB and diamagnetic background flows as well as time-evolving bootstrap current. The pellets are injected at different times in the inter-ELM phase with the pedestal build-up modelled via prescribed ad-hoc diffusion profiles, obtained from Ref.~\cite{Cathey2020}. This simplified approach allows to recover the experimentally observed lag-time for the first time in simulations and to investigate the plasma dynamics resulting from pellet injections at different phases of the build-up. 
Simulations with $0.8 \times 10^{20}$ deuterium atoms contained in the pellet (corresponding to an approximately two times larger pellet size in the experiment before the losses occurring in the guide tube), show a sharp transition of the energy losses between early ($t_\mathrm{inj.} \leq 10~\mathrm{ms}$) and later injection times ($t_\mathrm{inj.}\geq 12~\mathrm{ms}$), which correspond to different stages of pedestal build-up. Similarly, the incident power onto the divertor targets increases sharply when the pellet is injected at $12~\mathrm{ms}$. To make the transition comparable to the experiment (since the simulated pedestal build-up time scales might not be exactly comparable to the experiment depending on the chosen setup) the transition is characterized in pedestal parameters. The threshold for pellet ELM triggering with $0.8 \times 10^{20}$D pellet injection with $v_\mathrm{p}=560$ m/s is between a pedestal pressure of $p_\mathrm{ped} = 11.2$ and 11.6 kPa according to Figure~\ref{fig:pedestal_profiles_time_full} and Table~\ref{table:pedestal_evolve}.

To assess the impact of the pellet size on the simulated lag-time, also pellets with $0.4 \times 10^{20}$D atoms and $1.5 \times 10^{20}$D atoms were studied. The small pellet size is below the sizes experimentally accessible in ASDEX Upgrade, and the large pellet size corresponds approximately to the upper limit of pellet sizes experimentally accessible. With the small pellet, no ELM triggering was observed up to time point 12 ms, i.e., the small pellet never triggers an ELM in this work. The large pellet injection shows a transition between no-ELM response and ELM-triggering between injections at $6~\mathrm{ms}$ and $8~\mathrm{ms}$ which corresponds to $p_{ \mathrm{ped}}=9.6$ kPa and 10.6 kPa. Thus, we observe a dependency of the lag-time on the injected pellet size. All pellet-triggered ELMs correspond to a crash of the pedestal profiles within $\sim 0.4$ ms, independent of the pellet size or pellet injection time. A pronounced difference in the toroidal mode spectrum was observed between the no-ELM and ELM triggering responses. In the no-ELM response cases, the spectrum is significantly narrower than in the ELM-triggering cases. The $n=1$ component directly induced by the pellet is dominant in all cases. 

The pellet-triggered ELM cases show a pronounced toroidal asymmetry of the heat deposition consistent with simulations of DIII-D~\cite{Futatani2014} and JET~\cite{Futatani2019}, where self-consistent plasma flows and bootstrap current had not been taken into account. The heat deposition along the toroidal angle is observed to be largely independent from the pellet size which triggers the ELM crash. The footprint  of the heat flux onto the divertor target is thus characterized by the magnetic field configuration, which is determined by the ELM crash itself. Similarities and differences between spontaneous and pellet-triggered ELM crashes are beyond the scope of this work and are separately under investigation.

Finally, the dependence of the plasma dynamics onto the pellet injection velocity has been studied and it is found that the injection velocity plays an important role in the plasma dynamics. 
The lag-time analysis on the slower pellets injections (300 m/s) has been performed with the large pellet ($1.5 \times 10^{20}$D). It is observed that the slow injection velocity shows the transition from no-ELM to ELM triggering in earlier time.
The dependence of the injection velocity of $1.5 \times 10^{20}$D pellets injected at 8 ms has been investigated for 560 m/s, 300 m/s and 240 m/s. In case of high speed injection velocity of the pellet, the magnetic energies show stronger growth with respect to the cases of slow injection velocity. This is due to the deeper penetration of the pellet in the plasma (the amplitude of the perturbation is larger for higher $v_\mathrm{p}$). When the pellet injection is fast, the pellet rapidly reaches the inside of the pedestal where the plasma temperature is high. With the local high temperature, the pellet ablates quickly and creates a large density perturbation near the pedestal top, which excites the MHD modes causing the ELM crash. A velocity scan for a smaller pellet ($0.8 \times 10^{20}$D atoms) injected at 14 ms shows that with an injection velocity 800 m/s, the pellet reaches to the flux surface of $q=3$ and $q=2$ ($\Psi_N \approx 0.72$ and 0.51, respectively). This pellet reaching the core, does not only trigger an ELM crash, but also produces a large enough perturbation for giving rise to the growth of a $2/1$ neoclassical tearing mode at the $q=2$ rational surface. The energy loss caused by pellet-triggered ELMs observes a non-monotonic dependency to the injection velocity (with all other pellet parameters kept constant).
It is important to note that pellet size and velocity can typically not be modified independently in the experiment. Larger pellets are injected with lower velocities. Consequently, the two effects observed in our simulation would cancel to some extent in experiments: large pellets trigger easier, but the slower injection velocity acts in the opposite direction.

As already mentioned, the comparison of spontaneous and pellet-triggered plasmas is studied separately. Further work will attempt to demonstrate pellet ELM pacing over several ELM cycles based on a simulation setup like described in Ref.~\cite{Cathey2020} for spontaneous ELMs. 
Additional work based on different plasma parameters and several injections for ELM pacing will also be performed to clarify why the lag time shows dependencies on pellet size and velocity in our simulations, that were not observed experimentally in AUG-W.
Furthermore, pellet injection into ELM mitigated plasmas will be studied in the future to investigate the compatibility of fuelling pellets with ELM mitigation or suppression.

\bibliography{main}

\begin{thebibliography}{36}
\providecommand{\natexlab}[1]{#1}
\providecommand{\url}[1]{\texttt{#1}}
\expandafter\ifx\csname urlstyle\endcsname\relax
  \providecommand{\doi}[1]{doi: #1}\else
  \providecommand{\doi}{doi: \begingroup \urlstyle{rm}\Url}\fi

\bibitem[Loarte et~al.(2014)Loarte, Huijsmans, Futatani, Baylor, Evans, Orlov,
  Schmitz, Becoulet, Cahyna, Gribov, Kavin, Naik, Campbell, Casper, Daly,
  Frerichs, Kischner, Laengner, Lisgo, Pitts, Saibene, and Wingen]{Loarte2014}
A~Loarte, G~Huijsmans, S.~Futatani, L.R. Baylor, T.E. Evans, D.~M. Orlov,
  O.~Schmitz, M.~Becoulet, P.~Cahyna, Y.~Gribov, A.~Kavin, A.~Sashala Naik,
  D.J. Campbell, T.~Casper, E.~Daly, H.~Frerichs, A.~Kischner, R.~Laengner,
  S.~Lisgo, R.A. Pitts, G.~Saibene, and A.~Wingen.
\newblock Progress on the application of {ELM} control schemes to {ITER}
  scenarios from the non-active phase to {DT} operation.
\newblock \emph{Nuclear Fusion}, 54\penalty0 (3):\penalty0 033007, 2014.
\newblock URL
  \url{https://iopscience.iop.org/article/10.1088/0029-5515/54/3/033007/meta}.

\bibitem[Evans et~al.(2008)Evans, Fenstermacher, Moyer, Osborne, Watkins,
  Gohil, Joseph, Schaffer, Baylor, Bécoulet, Boedo, Burrell, deGrassie,
  Finken, Jernigan, Jakubowski, Lasnier, Lehnen, Leonard, Lonnroth, Nardon,
  Parail, Schmitz, Unterberg, and West]{Evans2008}
T.E. Evans, M.E. Fenstermacher, R.A. Moyer, T.H. Osborne, J.G. Watkins,
  P.~Gohil, I.~Joseph, M.J. Schaffer, L.R. Baylor, M.~Bécoulet, J.A. Boedo,
  K.H. Burrell, J.S. deGrassie, K.H. Finken, T.~Jernigan, M.W. Jakubowski, C.J.
  Lasnier, M.~Lehnen, A.W. Leonard, J.~Lonnroth, E.~Nardon, V.~Parail,
  O.~Schmitz, B.~Unterberg, and W.P. West.
\newblock Rmp {ELM} suppression in diii-d plasmas with iter similar shapes and
  collisionalities.
\newblock \emph{Nuclear Fusion}, 48\penalty0 (2):\penalty0 024002, 2008.
\newblock URL \url{http://stacks.iop.org/0029-5515/48/i=2/a=024002}.

\bibitem[Lang et~al.(2004)Lang, Conway, Eich, Fattorini, Gruber, Günter,
  Horton, Kalvin, Kallenbach, Kaufmann, Kocsis, Lorenz, Manso, Maraschek,
  Mertens, Neuhauser, Nunes, Schneider, Suttrop, Urano, and {the ASDEX Upgrade
  Team}]{Lang2004_1}
P.T. Lang, G.D. Conway, T.~Eich, L.~Fattorini, O.~Gruber, S.~Günter, L.D.
  Horton, S.~Kalvin, A.~Kallenbach, M.~Kaufmann, G.~Kocsis, A.~Lorenz, M.E.
  Manso, M.~Maraschek, V.~Mertens, J.~Neuhauser, I.~Nunes, W.~Schneider,
  W.~Suttrop, H.~Urano, and {the ASDEX Upgrade Team}.
\newblock {ELM} pace making and mitigation by pellet injection in {ASDEX
  Upgrade}.
\newblock \emph{Nuclear Fusion}, 44\penalty0 (5):\penalty0 665, 2004.
\newblock URL
  \url{https://iopscience.iop.org/article/10.1088/0029-5515/44/5/010/meta}.

\bibitem[Lang et~al.(2003)Lang, Neuhauser, Horton, Eich, Fattorini1, Fuchs,
  Gehre, Herrmann, Ignacz, Jakobi, Kalvin, Kaufmann, Kocsis, Kurzan, Maggi,
  Manso1, Maraschek, Mertens, Muck, Murmann, Neu, Nunes, Reich, Reich,
  Saarelma, Sandmann, Stober, Vogl, , and {the ASDEX Upgrade Team}]{Lang2003}
P.T. Lang, J.~Neuhauser, L.D. Horton, T.~Eich, L.~Fattorini1, J.C. Fuchs,
  O.~Gehre, A.~Herrmann, P.~Ignacz, M.~Jakobi, S.~Kalvin, M.~Kaufmann,
  G.~Kocsis, B.~Kurzan, C.~Maggi, M.E. Manso1, M.~Maraschek, V.~Mertens,
  A.~Muck, H.D. Murmann, R.~Neu, I.~Nunes, D.~Reich, M.~Reich, S.~Saarelma,
  W.~Sandmann, J.~Stober, U.~Vogl, , and {the ASDEX Upgrade Team}.
\newblock {ELM} frequency control by continuous small pellet injection in asdex
  upgrade.
\newblock \emph{Nuclear Fusion}, 43\penalty0 (10):\penalty0 1110, 2003.
\newblock URL
  \url{https://iopscience.iop.org/article/10.1088/0029-5515/43/10/012}.

\bibitem[Baylor et~al.(2013{\natexlab{a}})Baylor, Commaux, Jernigan, Meitner,
  Combs, Isler, Unterberg, Brooks, Evans, Leonard, Osborne, Parks, Snyder,
  Strait, Fenstermacher, Lasnier, Moyer, Loarte, Huijsmans, , and
  Futatani]{Baylor2013POP}
L.~R. Baylor, N.~Commaux, T.~C. Jernigan, S.~J. Meitner, S.~K. Combs, R.~C.
  Isler, E.~A. Unterberg, N.~H. Brooks, T.~E. Evans, A.~W. Leonard, T.~H.
  Osborne, P.~B. Parks, P.~B. Snyder, E.~J. Strait, M.~E. Fenstermacher, C.~J.
  Lasnier, R.~A. Moyer, A.~Loarte, G.~T.~A. Huijsmans, , and S.~Futatani.
\newblock Reduction of edge localized mode intensity on {DIII-D} by on-demand
  triggering with high frequency pellet injection and implications for {ITER}.
\newblock \emph{Physics of Plasmas}, 20\penalty0 (8):\penalty0 082513,
  2013{\natexlab{a}}.
\newblock URL \url{https://aip.scitation.org/doi/10.1063/1.4818772}.

\bibitem[Lang et~al.(2014)Lang, Burckhart, Bernert, Casali, Fischer, Kardaun,
  Kocsis, Maraschek, Mlynek, Plöckl, Reich, Ryter, Schweinzer, Sieglin,
  Suttrop, Szepesi, Tardini, Wolfrum, Zasche, and and]{Lang_2014}
P.T. Lang, A.~Burckhart, M.~Bernert, L.~Casali, R.~Fischer, O.~Kardaun,
  G.~Kocsis, M.~Maraschek, A.~Mlynek, B.~Plöckl, M.~Reich, F.~Ryter,
  J.~Schweinzer, B.~Sieglin, W.~Suttrop, T.~Szepesi, G.~Tardini, E.~Wolfrum,
  D.~Zasche, and H.~Zohm and.
\newblock {ELM} pacing and high-density operation using pellet injection in the
  {ASDEX} upgrade all-metal-wall tokamak.
\newblock \emph{Nuclear Fusion}, 54\penalty0 (8):\penalty0 083009, jun 2014.
\newblock \doi{10.1088/0029-5515/54/8/083009}.
\newblock URL \url{https://doi.org/10.1088%2F0029-5515%2F54%2F8%2F083009}.

\bibitem[Cathey et~al.(2020)Cathey, Hoelzl, Lackner, Huijsmans, Dunne, Wolfrum,
  Pamela, Orain, Günter, the {JOREK team}, the {ASDEX Upgrade team}, and the
  {EUROfusion MST1 team}]{Cathey2020}
A.~Cathey, M.~Hoelzl, K.~Lackner, G.T.A. Huijsmans, M.G. Dunne, E.~Wolfrum,
  S.J.P. Pamela, F.~Orain, S.~Günter, the {JOREK team}, the {ASDEX Upgrade
  team}, and the {EUROfusion MST1 team}.
\newblock Non-linear extended {MHD} simulations of {type-I} edge localised mode
  cycles in {ASDEX Upgrade} and their underlying triggering mechanism.
\newblock \emph{Nuclear Fusion}, 60:\penalty0 124007, 2020.
\newblock \doi{https://doi.org/10.1088/1741-4326/abbc87}.

\bibitem[{A. Herrmann (Guest editor)}(2003)]{Herrmann2003}
{A. Herrmann (Guest editor)}.
\newblock {Special Issue on ASDEX Upgrade}.
\newblock \emph{Fusion Science and Technology}, 44:\penalty0 1--747, 2003.

\bibitem[Lang et~al.(2007)Lang, Alper, Buttery, Gal, Hobirk, Neuhauser, Stamp,
  and {JET-EFDA contributors}]{Lang2007}
P.T. Lang, B.~Alper, R.~Buttery, K.~Gal, J.~Hobirk, J.~Neuhauser, M.~Stamp, and
  {JET-EFDA contributors}.
\newblock {ELM} triggering by local pellet perturbations in {type-I ELMy
  H-mode} plasma at {JET}.
\newblock \emph{Nuclear Fusion}, 47\penalty0 (8):\penalty0 754, 2007.
\newblock URL
  \url{https://iopscience.iop.org/article/10.1088/0029-5515/47/8/005/pdf}.

\bibitem[Lang et~al.(2006)Lang, Gal, Hobirk, Kalvin, Kocsis, Mertens,
  Neuhauser, Maraschek, Suttrop, Veres, and {the ASDEX Upgrade Team}]{Lang2006}
P~T Lang, K~Gal, J~Hobirk, S~Kalvin, G~Kocsis, V~Mertens, J~Neuhauser,
  M~Maraschek, W~Suttrop, G~Veres, and {the ASDEX Upgrade Team}.
\newblock Investigations on the {ELM} cycle by local {3D} perturbation
  experiments.
\newblock \emph{Plasma Physics and Controlled Fusion}, 48\penalty0
  (5A):\penalty0 A141, 2006.
\newblock URL
  \url{https://iopscience.iop.org/article/10.1088/0741-3335/48/5A/S13/pdf}.

\bibitem[Baylor et~al.(2013{\natexlab{b}})Baylor, Commaux, Jernigan, Brooks,
  Combs, Evans, Fenstermacher, Isler, Lasnier, Meitner, Moyer, Osborne, Parks,
  Snyder, Strait, Unterberg, , and Loarte]{Baylor2013PRL}
L.~R. Baylor, N.~Commaux, T.~C. Jernigan, N.~H. Brooks, S.~K. Combs, T.~E.
  Evans, M.~E. Fenstermacher, R.~C. Isler, C.~J. Lasnier, S.~J. Meitner, R.~A.
  Moyer, T.~H. Osborne, P.~B. Parks, P.~B. Snyder, E.~J. Strait, E.~A.
  Unterberg, , and A.~Loarte.
\newblock Reduction of edge-localized mode intensity using high-repetition-rate
  pellet injection in tokamak {H-Mode} plasmas.
\newblock \emph{Physical Review Letters}, 110:\penalty0 245001,
  2013{\natexlab{b}}.
\newblock URL \url{https://doi.org/10.1103/PhysRevLett.110.245001}.

\bibitem[Romanelli et~al.(2009)Romanelli, Kamendje, and {on behalf of JET-EFDA
  Contributors}]{Romanelli2009}
F.~Romanelli, R.~Kamendje, and {on behalf of JET-EFDA Contributors}.
\newblock Overview of {JET} results.
\newblock \emph{Nuclear Fusion}, 49\penalty0 (10):\penalty0 104006, 2009.
\newblock URL
  \url{https://iopscience.iop.org/article/10.1088/0029-5515/49/10/104006/meta}.

\bibitem[Kocsis et~al.(2007)Kocsis, Kálvin, Lang, Maraschek, Neuhauser,
  Schneider, Szepesi, and {the ASDEX Upgrade Team}]{Kocsis2007}
G.~Kocsis, S.~Kálvin, P.T. Lang, M.~Maraschek, J.~Neuhauser, W.~Schneider,
  T.~Szepesi, and {the ASDEX Upgrade Team}.
\newblock Spatio-temporal investigations on the triggering of pellet induced
  {ELMs}.
\newblock \emph{Nuclear Fusion}, 47\penalty0 (9):\penalty0 1166, 2007.
\newblock URL
  \url{https://iopscience.iop.org/article/10.1088/0029-5515/47/9/013/pdf}.

\bibitem[Lang et~al.(2008)Lang, Lackner, Maraschek, Alper, Belonohy, Gál,
  Hobirk, Kallenbach, Kálvin, Kocsis, von Thun, Suttrop, Szepesi, Wenninger,
  Zohm, {the ASDEX Upgrade Team}, and {JET-EFDA contributors}]{Lang2008}
P.T. Lang, K.~Lackner, M.~Maraschek, B.~Alper, E.~Belonohy, K.~Gál, J.~Hobirk,
  A.~Kallenbach, S.~Kálvin, G.~Kocsis, C.P.~Perez von Thun, W.~Suttrop,
  T.~Szepesi, R.~Wenninger, H.~Zohm, {the ASDEX Upgrade Team}, and {JET-EFDA
  contributors}.
\newblock Investigation of pellet-triggered {MHD} events in {ASDEX Upgrade} and
  {JET}.
\newblock \emph{Nuclear Fusion}, 48\penalty0 (9):\penalty0 095007, 2008.
\newblock URL \url{http://stacks.iop.org/0029-5515/48/i=9/a=095007}.

\bibitem[Lang et~al.(2015)Lang, Meyer, Birkenmeier, Burckhart, Carvalho,
  Delabie, Frassinetti, Huijsmans, G~Kocsi~and, Maggi, Maraschek, Ploeckl,
  Rimini, Ryter, Saarelma, Szepesi, Wolfrum, {ASDEX Upgrade Team}, and {JET
  Contributors}]{Lang2015}
P~T Lang, H~Meyer, G~Birkenmeier, A~Burckhart, I~S Carvalho, E~Delabie,
  L~Frassinetti, G~Huijsmans, A~Loarte G~Kocsi~and, C~F Maggi, M~Maraschek,
  B~Ploeckl, F~Rimini, F~Ryter, S~Saarelma, T~Szepesi, E~Wolfrum, {ASDEX
  Upgrade Team}, and {JET Contributors}.
\newblock {ELM} control at the {L-H} transition by means of pellet pacing in
  the asdex upgrade and {JET} all-metal-wall tokamaks.
\newblock \emph{Plasma Physics and Controlled Fusion}, 57\penalty0
  (4):\penalty0 045011, 2015.
\newblock URL
  \url{https://iopscience.iop.org/article/10.1088/0741-3335/57/4/045011/meta}.

\bibitem[Poli et~al.(2010)Poli, Lang, Sharapov, Alper, Koslowski, and {JET-EFDA
  contributors}]{Poli2010}
F.M. Poli, P.T. Lang, S.E. Sharapov, B.~Alper, H.R. Koslowski, and {JET-EFDA
  contributors}.
\newblock Spectra of magnetic perturbations triggered by pellets in jet
  plasmas.
\newblock \emph{Nuclear Fusion}, 50:\penalty0 025004, 2010.
\newblock \doi{https://doi.org/10.1088/0029-5515/50/2/025004}.

\bibitem[Szepesi et~al.(2009)Szepesi, K\'{a}lvin, Kocsis, Lackner, Maraschek,
  Pokol, Por, and {ASDEX Upgrade Team}]{Szepesi2009}
T~Szepesi, S~K\'{a}lvin, G~Kocsis, K~Lackner, P~T Lang~M Maraschek, G~Pokol,
  G~Por, and {ASDEX Upgrade Team}.
\newblock Plasma physics and controlled fusion investigation of pellet-driven
  magnetic perturbations in different tokamak scenarios.
\newblock \emph{Plasma Phys. Control. Fusion}, 51:\penalty0 125002, 2009.
\newblock \doi{https://doi.org/10.1088/0741-3335/51/12/125002}.

\bibitem[Sovinec et~al.(2004)Sovinec, Glasser, Gianakon, Barnes, Nebel, Kruger,
  Plimpton, Tarditi, Chu, and {the NIMROD Team}]{Sovinec2004}
C.R. Sovinec, A.H. Glasser, T.A. Gianakon, D.C. Barnes, R.A. Nebel, S.E.
  Kruger, S.J. Plimpton, A.~Tarditi, M.S. Chu, and {the NIMROD Team}.
\newblock Nonlinear magnetohydrodynamics with high-order finite elements.
\newblock \emph{J. Comp. Phys.}, 195:\penalty0 355, 2004.

\bibitem[Ferraro et~al.(2010)Ferraro, Jardin, and Snyder]{Ferraro2010}
N.~M. Ferraro, S.~C. Jardin, and P.~B. Snyder.
\newblock Ideal and resistive edge stability calculations with m3d-c1.
\newblock \emph{Physics of Plasmas}, 17:\penalty0 102508, 2010.

\bibitem[Dudson et~al.(2009)Dudson, Umansky, Xu, Snyder, and
  Wilson]{Dudson2009}
B.D. Dudson, M.V. Umansky, X.Q. Xu, P.B. Snyder, and H.R. Wilson.
\newblock Bout++: A framework for parallel plasma fluid simulations.
\newblock \emph{Computer Physics Communications}, 180\penalty0 (9):\penalty0
  1467 -- 1480, 2009.
\newblock ISSN 0010-4655.
\newblock \doi{10.1016/j.cpc.2009.03.008}.

\bibitem[Huysmans and Czarny(2007)]{Huysmans2007}
G.T.A. Huysmans and O.~Czarny.
\newblock {MHD} stability in x-point geometry: simulation of {ELMs}.
\newblock \emph{Nuclear Fusion}, 47\penalty0 (7):\penalty0 659, 2007.
\newblock \doi{10.1088/0029-5515/47/7/016}.

\bibitem[B\'ecoulet et~al.(2014)B\'ecoulet, Orain, Huijsmans, Pamela, Cahyna,
  Hoelzl, Garbet, Franck, Sonnendr\"ucker, Dif-Pradalier, Passeron, Latu,
  Morales, Nardon, Fil, Nkonga, Ratnani, and Grandgirard]{Becoulet2014}
M.~B\'ecoulet, F.~Orain, G.~T.~A. Huijsmans, S.~Pamela, P.~Cahyna, M.~Hoelzl,
  X.~Garbet, E.~Franck, E.~Sonnendr\"ucker, G.~Dif-Pradalier, C.~Passeron,
  G.~Latu, J.~Morales, E.~Nardon, A.~Fil, B.~Nkonga, A.~Ratnani, and
  V.~Grandgirard.
\newblock Mechanism of edge localized mode mitigation by resonant magnetic
  perturbations.
\newblock \emph{Phys. Rev. Lett.}, 113:\penalty0 115001, Sep 2014.
\newblock \doi{10.1103/PhysRevLett.113.115001}.

\bibitem[Hoelzl et~al.(2020)Hoelzl, Hu, Nardon, and Huijsmans]{Hoelzl2020A}
M.~Hoelzl, D.~Hu, E.~Nardon, and G.~T.~A. Huijsmans.
\newblock First predictive simulations for deuterium shattered pellet injection
  in {ASDEX Upgrade}.
\newblock \emph{Physics of Plasmas}, 27\penalty0 (2):\penalty0 022510, 2020.
\newblock \doi{10.1063/1.5133099}.

\bibitem[Orain et~al.(2013)Orain, B\'ecoulet, Dif-Pradalier, Huijsmans, Pamela,
  Nardon, Passeron, Latu, Grandgirard, Fil, Ratnani, Chapman, Kirk, Thornton,
  Hoelzl, and Cahyna]{Orain2013}
F.~Orain, M.~B\'ecoulet, G.~Dif-Pradalier, G.~Huijsmans, S.~Pamela, E.~Nardon,
  C.~Passeron, G.~Latu, V.~Grandgirard, A.~Fil, A.~Ratnani, I.~Chapman,
  A.~Kirk, A.~Thornton, M.~Hoelzl, and P.~Cahyna.
\newblock Non-linear magnetohydrodynamic modeling of plasma response to
  resonant magnetic perturbations.
\newblock \emph{Physics of Plasmas}, 20\penalty0 (10):\penalty0 102510, 2013.
\newblock \doi{10.1063/1.4824820}.

\bibitem[Artola et~al.(2018)Artola, Huijsmans, Hoelzl, Beyer, Loarte, and
  Gribov]{Artola2018}
F.J. Artola, G.T.A. Huijsmans, M.~Hoelzl, P.~Beyer, A.~Loarte, and Y.~Gribov.
\newblock Non-linear magnetohydrodynamic simulations of edge localised mode
  triggering via vertical position oscillations in {ITER}.
\newblock \emph{Nuclear Fusion}, 58\penalty0 (9):\penalty0 096018, jul 2018.
\newblock \doi{10.1088/1741-4326/aace0e}.

\bibitem[Liu et~al.(2015)Liu, Huijsmans, Loarte, Garofalo, Solomon, Snyder,
  Hoelzl, and Zeng]{LiuF2015}
F.~Liu, G.T.A. Huijsmans, A.~Loarte, A.M. Garofalo, W.M. Solomon, P.B. Snyder,
  M.~Hoelzl, and L.~Zeng.
\newblock Nonlinear {MHD} simulations of {Quiescent H-mode} plasmas in
  {DIII-D}.
\newblock \emph{Nuclear Fusion}, 55\penalty0 (11):\penalty0 113002, 2015.
\newblock URL \url{http://stacks.iop.org/0029-5515/55/i=11/a=113002}.

\bibitem[Futatani et~al.(2014)Futatani, Huijsmans, Loarte, Baylor, Commaux,
  Jernigan, Fenstermacher, Lasnier, Osborne, and Pegourié]{Futatani2014}
S.~Futatani, G.~Huijsmans, A.~Loarte, L.R. Baylor, N.~Commaux, T.C. Jernigan,
  M.E. Fenstermacher, C.~Lasnier, T.H. Osborne, and B.~Pegourié.
\newblock Non-linear {MHD} modelling of {ELM} triggering by pellet injection in
  {DIII-D} and implications for {ITER}.
\newblock \emph{Nuclear Fusion}, 54\penalty0 (7):\penalty0 073008, 2014.
\newblock URL \url{http://stacks.iop.org/0029-5515/54/i=7/a=073008}.

\bibitem[Diem et~al.(2019)Diem, Baylor, Ferraro, Lyons, Shiraki, and
  Wilcox]{Diem2019}
S.J. Diem, L.R. Baylor, N.M. Ferraro, B.C. Lyons, D.~Shiraki, and R.S. Wilcox.
\newblock Utilizing m3d-c1 to understand triggering of elms in pellet pacing
  experiments in diii-d iter-like plasmas.
\newblock \emph{Proceedings of the 46th EPS Conference on Plasma Physics}, page
  P1.1055, 2019.

\bibitem[Fil et~al.(2017)Fil, Kolemen, A.Bortolon, N.Ferraro, S.Jardin,
  P.B.Parks, R.Lunsford, and R.Maingi]{Fil2017}
A.~Fil, E.~Kolemen, A.Bortolon, N.Ferraro, S.Jardin, P.B.Parks, R.Lunsford, and
  R.Maingi.
\newblock Modeling of lithium granule injection in nstx with m3d-c1.
\newblock \emph{Nuclear Materials and Energy}, 12:\penalty0 1094, 2017.
\newblock URL
  \url{https://www.sciencedirect.com/science/article/pii/S2352179116302708#bib0001}.

\bibitem[Huysmans et~al.(2009)Huysmans, Pamela, van~der Plas, and
  Ramet]{Huysmans2009}
G~T~A Huysmans, S~Pamela, E~van~der Plas, and P~Ramet.
\newblock Non-linear {MHD} simulations of edge localized modes ({ELMs}).
\newblock \emph{Plasma Physics and Controlled Fusion}, 51\penalty0
  (12):\penalty0 124012, 2009.
\newblock URL \url{http://stacks.iop.org/0741-3335/51/i=12/a=124012}.

\bibitem[Futatani et~al.(2019)Futatani, Pamela, Garzotti, Huijsmans, Hoelzl,
  Frigione, Lennholm, and and]{Futatani2019}
S.~Futatani, S.~Pamela, L.~Garzotti, G.T.A. Huijsmans, M.~Hoelzl, D.~Frigione,
  M.~Lennholm, and and.
\newblock Non-linear magnetohydrodynamic simulations of pellet triggered
  edge-localized modes in {JET}.
\newblock \emph{Nuclear Fusion}, 60\penalty0 (2):\penalty0 026003, dec 2019.
\newblock \doi{10.1088/1741-4326/ab56c7}.

\bibitem[Frigione et~al.(2015)Frigione, L.Garzotti, M.Lennholm, B.Alper,
  G.Artaserse, P.Bennett, E.Giovannozzi, T.Eich, G.Kocsis, P.T.Lang,
  G.Maddaluno, R.Mooney, M.Rack, G.Sips, G.Tvalashvili, B.Viola, D.Wilkes, and
  {JET-EFDA Contributors}]{Frigione2015}
D.~Frigione, L.Garzotti, M.Lennholm, B.Alper, G.Artaserse, P.Bennett,
  E.Giovannozzi, T.Eich, G.Kocsis, P.T.Lang, G.Maddaluno, R.Mooney, M.Rack,
  G.Sips, G.Tvalashvili, B.Viola, D.Wilkes, and {JET-EFDA Contributors}.
\newblock Divertor load footprint of {ELMs} in pellet triggering and pacing
  experiments at {JET}.
\newblock \emph{Journal of Nuclear Materials}, 463:\penalty0 714, 2015.
\newblock URL
  \url{https://www.sciencedirect.com/science/article/abs/pii/S002231151500063X}.

\bibitem[Wenninger et~al.(2011)Wenninger, Eich, Huysmans, Lang, Devaux,
  Jachmich, Köchl, and {JET EFDA Contributors}]{Wenninger2011}
R~P Wenninger, T~H Eich, G~T~A Huysmans, P~T Lang, S~Devaux, S~Jachmich,
  F~Köchl, and {JET EFDA Contributors}.
\newblock Scrape-off layer heat transport and divertor power deposition of
  pellet-induced edge localized modes.
\newblock \emph{Plasma Physics and Controlled Fusion}, 53\penalty0
  (10):\penalty0 105002, aug 2011.
\newblock \doi{10.1088/0741-3335/53/10/105002}.

\bibitem[Meyer et~al.(2019)Meyer, Angioni, Albert, Arden, Parra, Asunta,
  de~Baar, Balden, Bandaru, Behler, Bergmann, Bernardo, Bernert, Biancalani,
  Bilato, Birkenmeier, Blanken, Bobkov, Bock, Bolzonella, Bortolon, Böswirth,
  Bottereau, Bottino, van~den Brand, Brezinsek, Brida, Brochard, Bruhn,
  Buchanan, Buhler, Burckhart, Camenen, Carlton, Carr, Carralero, Castaldo,
  Cavedon, Cazzaniga, Ceccuzzi, Challis, Chankin, Chapman, Cianfarani, Clairet,
  Coda, Coelho, Coenen, Colas, Conway, Costea, Coster, Cote, Creely, Croci,
  Cseh, Czarnecka, Cziegler, D'Arcangelo, David, Day, Delogu, de~Marn{\'{e}},
  Denk, Denner, Dibon, Siena, Douai, Drenik, Drube, Dunne, Duval, Dux, Eich,
  Elgeti, Engelhardt, Erdös, Erofeev, Esposito, Fable, Faitsch, Fantz, Faugel,
  Faust, Felici, Ferreira, Fietz, Figuereido, Fischer, Ford, Frassinetti,
  Freethy, Fröschle, Fuchert, Fuchs, Fünfgelder, Galazka, Galdon-Quiroga,
  Gallo, Gao, Garavaglia, Garcia-Carrasco, Garcia-Mu{\~{n}}oz, Geiger,
  Giannone, Gil, Giovannozzi, Gleason-Gonz{\'{a}}lez, Glöggler, Gobbin,
  Görler, Ortiz, Martin, Goodman, Gorini, Gradic, Gräter, Granucci, Greuner,
  Griener, Groth, Gude, Günter, Guimarais, Haas, Hakola, Ham, Happel, den
  Harder, Harrer, Harrison, Hauer, Hayward-Schneider, Hegna, Heinemann,
  Heinzel, Hellsten, Henderson, Hennequin, Herrmann, Heyn, Heyn, Hitzler,
  Hobirk, Höfler, Hölzl, Höschen, Holm, Hopf, Hornsby, Horvath, Houben,
  Huber, Igochine, Ilkei, Ivanova-Stanik, Jacob, Jacobsen, Janky, van Vuuren,
  Jardin, Jaulmes, Jenko, Jensen, Joffrin, Käsemann, Kallenbach, K{\'{a}}lvin,
  Kantor, Kappatou, Kardaun, Karhunen, Kasilov, Kazakov, Kernbichler, Kirk,
  Hansen, Klevarova, Kocsis, Köhn, Koubiti, Krieger, Krivska, Krämer-Flecken,
  Kudlacek, Kurki-Suonio, Kurzan, Labit, Lackner, Laggner, Lang, Lauber,
  Lebschy, Leuthold, Li, Linder, Lipschultz, Liu, Liu, Lohs, Lu,
  di~Cortemiglia, Luhmann, Lunsford, Lunt, Lyssoivan, Maceina, Madsen,
  Maggiora, Maier, Maj, Mailloux, Maingi, Maljaars, Manas, Mancini, Manhard,
  Manso, Mantica, Mantsinen, Manz, Maraschek, Martens, Martin, Marrelli,
  Martitsch, Mayer, Mazon, McCarthy, McDermott, Meister, Medvedeva, Merkel,
  Merle, Mertens, Meshcheriakov, Meyer, Miettunen, Milanesio, Mink, Mlynek,
  Monaco, Moon, Nabais, Nemes-Czopf, Neu, Neu, Nielsen, Nielsen, Nikolaeva,
  Nocente, Noterdaeme, Novikau, Nowak, Oberkofler, Oberparleiter, Ochoukov,
  Odstrcil, Olsen, Orain, Palermo, Pan, Papp, Perez, Pau, Pautasso, Penzel,
  Petersson, Acosta, Piovesan, Piron, Pitts, Plank, Plaum, Ploeckl, Plyusnin,
  Pokol, Poli, Porte, Potzel, Prisiazhniuk, Pütterich, Ramisch, Rasmussen,
  Ratt{\'{a}}, Ratynskaia, Raupp, Ravera, R{\'{e}}fy, Reich, Reimold, Reiser,
  Ribeiro, Riesch, Riedl, Rittich, Rivero-Rodriguez, Rocchi, Rodriguez-Ramos,
  Rohde, Ross, Rott, Rubel, Ryan, Ryter, Saarelma, Salewski, Salmi,
  Sanchis-Sanchez, Santos, Sauter, Scarabosio, Schall, Schmid, Schmitz,
  Schneider, Schrittwieser, Schubert, Schwarz-Selinger, Schweinzer, Scott,
  Sehmer, Seliunin, Sertoli, Shabbir, Shalpegin, Shao, Sharapov, Sias,
  Siccinio, Sieglin, Sigalov, Silva, Silva, Silvagni, Simon, Simpson,
  Smigelskis, Snicker, Sommariva, Sozzi, Spolaore, Stegmeir, Stejner, Stober,
  Stroth, Strumberger, Suarez, Sun, Suttrop, Sytova, Szepesi, T{\'{a}}l, Tala,
  Tardini, Tardocchi, Teschke, Terranova, Tierens, Thor{\'{e}}n, Told, Tolias,
  Tudisco, Treutterer, Trier, Tripsk{\'{y}}, Valisa, Valovic, Vanovac, van
  Vugt, Varoutis, Verdoolaege, Vianello, Vicente, Vierle, Viezzer, von
  Toussaint, Wagner, Wang, Wang, Weiland, White, Wiesen, Willensdorfer,
  Wiringer, Wischmeier, Wolf, Wolfrum, Xiang, Yang, Yang, Yu, Zag{\'{o}}rski,
  Zammuto, Zhang, van Zeeland, Zehetbauer, Zilker, Zoletnik, and
  Zohm]{Meyer2019}
H.~Meyer, C.~Angioni, C.G. Albert, N.~Arden, R.~Arredondo Parra, O.~Asunta,
  M.~de~Baar, M.~Balden, V.~Bandaru, K.~Behler, A.~Bergmann, J.~Bernardo,
  M.~Bernert, A.~Biancalani, R.~Bilato, G.~Birkenmeier, T.C. Blanken,
  V.~Bobkov, A.~Bock, T.~Bolzonella, A.~Bortolon, B.~Böswirth, C.~Bottereau,
  A.~Bottino, H.~van~den Brand, S.~Brezinsek, D.~Brida, F.~Brochard, C.~Bruhn,
  J.~Buchanan, A.~Buhler, A.~Burckhart, Y.~Camenen, D.~Carlton, M.~Carr,
  D.~Carralero, C.~Castaldo, M.~Cavedon, C.~Cazzaniga, S.~Ceccuzzi, C.~Challis,
  A.~Chankin, S.~Chapman, C.~Cianfarani, F.~Clairet, S.~Coda, R.~Coelho, J.W.
  Coenen, L.~Colas, G.D. Conway, S.~Costea, D.P. Coster, T.B. Cote, A.~Creely,
  G.~Croci, G.~Cseh, A.~Czarnecka, I.~Cziegler, O.~D'Arcangelo, P.~David,
  C.~Day, R.~Delogu, P.~de~Marn{\'{e}}, S.S. Denk, P.~Denner, M.~Dibon, A.~Di
  Siena, D.~Douai, A.~Drenik, R.~Drube, M.~Dunne, B.P. Duval, R.~Dux, T.~Eich,
  S.~Elgeti, K.~Engelhardt, B.~Erdös, I.~Erofeev, B.~Esposito, E.~Fable,
  M.~Faitsch, U.~Fantz, H.~Faugel, I.~Faust, F.~Felici, J.~Ferreira, S.~Fietz,
  A.~Figuereido, R.~Fischer, O.~Ford, L.~Frassinetti, S.~Freethy, M.~Fröschle,
  G.~Fuchert, J.C. Fuchs, H.~Fünfgelder, K.~Galazka, J.~Galdon-Quiroga,
  A.~Gallo, Y.~Gao, S.~Garavaglia, A.~Garcia-Carrasco, M.~Garcia-Mu{\~{n}}oz,
  B.~Geiger, L.~Giannone, L.~Gil, E.~Giovannozzi, C.~Gleason-Gonz{\'{a}}lez,
  S.~Glöggler, M.~Gobbin, T.~Görler, I.~Gomez Ortiz, J.~Gonzalez Martin,
  T.~Goodman, G.~Gorini, D.~Gradic, A.~Gräter, G.~Granucci, H.~Greuner,
  M.~Griener, M.~Groth, A.~Gude, S.~Günter, L.~Guimarais, G.~Haas, A.H.
  Hakola, C.~Ham, T.~Happel, N.~den Harder, G.F. Harrer, J.~Harrison, V.~Hauer,
  T.~Hayward-Schneider, C.C. Hegna, B.~Heinemann, S.~Heinzel, T.~Hellsten,
  S.~Henderson, P.~Hennequin, A.~Herrmann, M.F. Heyn, E.~Heyn, F.~Hitzler,
  J.~Hobirk, K.~Höfler, M.~Hölzl, T.~Höschen, J.H. Holm, C.~Hopf, W.A.
  Hornsby, L.~Horvath, A.~Houben, A.~Huber, V.~Igochine, T.~Ilkei,
  I.~Ivanova-Stanik, W.~Jacob, A.S. Jacobsen, F.~Janky, A.~Jansen van Vuuren,
  A.~Jardin, F.~Jaulmes, F.~Jenko, T.~Jensen, E.~Joffrin, C.-P. Käsemann,
  A.~Kallenbach, S.~K{\'{a}}lvin, M.~Kantor, A.~Kappatou, O.~Kardaun,
  J.~Karhunen, S.~Kasilov, Y.~Kazakov, W.~Kernbichler, A.~Kirk, S.~Kjer Hansen,
  V.~Klevarova, G.~Kocsis, A.~Köhn, M.~Koubiti, K.~Krieger, A.~Krivska,
  A.~Krämer-Flecken, O.~Kudlacek, T.~Kurki-Suonio, B.~Kurzan, B.~Labit,
  K.~Lackner, F.~Laggner, P.T. Lang, P.~Lauber, A.~Lebschy, N.~Leuthold, M.~Li,
  O.~Linder, B.~Lipschultz, F.~Liu, Y.~Liu, A.~Lohs, Z.~Lu, T.~Luda
  di~Cortemiglia, N.C. Luhmann, R.~Lunsford, T.~Lunt, A.~Lyssoivan, T.~Maceina,
  J.~Madsen, R.~Maggiora, H.~Maier, O.~Maj, J.~Mailloux, R.~Maingi,
  E.~Maljaars, P.~Manas, A.~Mancini, A.~Manhard, M.-E. Manso, P.~Mantica,
  M.~Mantsinen, P.~Manz, M.~Maraschek, C.~Martens, P.~Martin, L.~Marrelli,
  A.~Martitsch, M.~Mayer, D.~Mazon, P.J. McCarthy, R.~McDermott, H.~Meister,
  A.~Medvedeva, R.~Merkel, A.~Merle, V.~Mertens, D.~Meshcheriakov, O.~Meyer,
  J.~Miettunen, D.~Milanesio, F.~Mink, A.~Mlynek, F.~Monaco, C.~Moon,
  F.~Nabais, A.~Nemes-Czopf, G.~Neu, R.~Neu, A.H. Nielsen, S.K. Nielsen,
  V.~Nikolaeva, M.~Nocente, J.-M. Noterdaeme, I.~Novikau, S.~Nowak,
  M.~Oberkofler, M.~Oberparleiter, R.~Ochoukov, T.~Odstrcil, J.~Olsen,
  F.~Orain, F.~Palermo, O.~Pan, G.~Papp, I.~Paradela Perez, A.~Pau,
  G.~Pautasso, F.~Penzel, P.~Petersson, J.~Pinz{\'{o}}n Acosta, P.~Piovesan,
  C.~Piron, R.~Pitts, U.~Plank, B.~Plaum, B.~Ploeckl, V.~Plyusnin, G.~Pokol,
  E.~Poli, L.~Porte, S.~Potzel, D.~Prisiazhniuk, T.~Pütterich, M.~Ramisch,
  J.~Rasmussen, G.A. Ratt{\'{a}}, S.~Ratynskaia, G.~Raupp, G.L. Ravera,
  D.~R{\'{e}}fy, M.~Reich, F.~Reimold, D.~Reiser, T.~Ribeiro, J.~Riesch,
  R.~Riedl, D.~Rittich, J.F. Rivero-Rodriguez, G.~Rocchi, M.~Rodriguez-Ramos,
  V.~Rohde, A.~Ross, M.~Rott, M.~Rubel, D.~Ryan, F.~Ryter, S.~Saarelma,
  M.~Salewski, A.~Salmi, L.~Sanchis-Sanchez, J.~Santos, O.~Sauter,
  A.~Scarabosio, G.~Schall, K.~Schmid, O.~Schmitz, P.A. Schneider,
  R.~Schrittwieser, M.~Schubert, T.~Schwarz-Selinger, J.~Schweinzer, B.~Scott,
  T.~Sehmer, E.~Seliunin, M.~Sertoli, A.~Shabbir, A.~Shalpegin, L.~Shao,
  S.~Sharapov, G.~Sias, M.~Siccinio, B.~Sieglin, A.~Sigalov, A.~Silva,
  C.~Silva, D.~Silvagni, P.~Simon, J.~Simpson, E.~Smigelskis, A.~Snicker,
  C.~Sommariva, C.~Sozzi, M.~Spolaore, A.~Stegmeir, M.~Stejner, J.~Stober,
  U.~Stroth, E.~Strumberger, G.~Suarez, H.-J. Sun, W.~Suttrop, E.~Sytova,
  T.~Szepesi, B.~T{\'{a}}l, T.~Tala, G.~Tardini, M.~Tardocchi, M.~Teschke,
  D.~Terranova, W.~Tierens, E.~Thor{\'{e}}n, D.~Told, P.~Tolias, O.~Tudisco,
  W.~Treutterer, E.~Trier, M.~Tripsk{\'{y}}, M.~Valisa, M.~Valovic, B.~Vanovac,
  D.~van Vugt, S.~Varoutis, G.~Verdoolaege, N.~Vianello, J.~Vicente, T.~Vierle,
  E.~Viezzer, U.~von Toussaint, D.~Wagner, N.~Wang, X.~Wang, M.~Weiland, A.E.
  White, S.~Wiesen, M.~Willensdorfer, B.~Wiringer, M.~Wischmeier, R.~Wolf,
  E.~Wolfrum, L.~Xiang, Q.~Yang, Z.~Yang, Q.~Yu, R.~Zag{\'{o}}rski, I.~Zammuto,
  W.~Zhang, M.~van Zeeland, T.~Zehetbauer, M.~Zilker, S.~Zoletnik, and H.~Zohm.
\newblock Overview of physics studies on {ASDEX} {Upgrade}.
\newblock \emph{Nuclear Fusion}, 59\penalty0 (11):\penalty0 112014, jul 2019.
\newblock \doi{10.1088/1741-4326/ab18b8}.

\bibitem[Czarny and Huysmans(2008)]{Czarny2008}
Olivier Czarny and Guido Huysmans.
\newblock Bezier surfaces and finite elements for {MHD} simulations.
\newblock \emph{Journal of Computational Physics}, 227\penalty0 (16):\penalty0
  7423 -- 7445, 2008.
\newblock ISSN 0021-9991.
\newblock \doi{10.1016/j.jcp.2008.04.001}.

\bibitem[Belonohy et~al.(2008)Belonohy, Kardaun, Fehér, Gál, Kálvin, Kocsis,
  Lackner, Lang, Neuhauser, and the {ASDEX Upgrade Team}]{Belonohy2008}
E.~Belonohy, O.J.W.F. Kardaun, T.~Fehér, K.~Gál, S.~Kálvin, G.~Kocsis,
  K.~Lackner, P.T. Lang, J.~Neuhauser, and the {ASDEX Upgrade Team}.
\newblock A high field side pellet penetration depth scaling derived for asdex
  upgrade.
\newblock \emph{Nuclear Fusion}, 48:\penalty0 065009, 2008.
\newblock
  \doi{https://iopscience.iop.org/article/10.1088/0029-5515/48/6/065009/pdf}.

\end{thebibliography}

\section*{Acknowledgements}

This work has been carried out within the framework of the EUROfusion Consortium and has received funding from the Euratom research and training program 2014-2018 and 2019-2020 under grant agreement No 633053. The views and opinions expressed herein do not necessarily reflect those of the European Commission. In particular, contributions by EUROfusion work packages Enabling Research (EnR) and Medium Size Tokamaks (MST) is acknowledged. The author thankfully acknowledges the computer resources of PRACE (Partnership for Advanced Computing in Europe) and RES (Spanish Supercomputing Network) at MareNostrum and the technical support provided by Barcelona Supercomputing Center (BSC), and the support from Marconi-Fusion, the High Performance Computer at the CINECA headquarters in Bologna (Italy) for its provision of supercomputer resources. 
The work of S.F. is supported by a Ram\'{o}n y Cajal grant from the Spanish Ministry of Economy and Competitiveness (RYC-2014-15206).

\end{document}